\documentclass[twocolumn,twocolappendix]{aastex63}

\usepackage{newtxtext,newtxmath}

\usepackage[T1]{fontenc}
\usepackage{ae,aecompl}
\usepackage{comment}
\usepackage{multirow}

\usepackage{graphicx,tabularx}	
\usepackage{amsmath}	
\usepackage{amssymb}	
\usepackage{booktabs}
\usepackage[inline,shortlabels]{enumitem}
\setlist{itemjoin* = { and\enspace}}

\received{09-Nov-2020}
\revised{21-Dec-2020}
\accepted{07-Jan-2021}

\shorttitle{Antisymmetric H{\small I}-CO Cross-correlation}
\shortauthors{Zhou, Tan \& Mao}
\graphicspath{{./}}

\begin{document}

\title{Antisymmetric Cross-correlation between H{\small I} and CO Line Intensity Maps as a New Probe of Cosmic Reionization}

\correspondingauthor{Yi Mao} \email{ymao@tsinghua.edu.cn}

\author[0000-0002-2744-0618]{Meng Zhou} 
\affiliation{Department of Astronomy, Tsinghua University, Beijing 100084, China}

\author[0000-0001-6161-7037]{Jianrong Tan} 
\affiliation{Department of Astronomy, Tsinghua University, Beijing 100084, China}
\affiliation{Department of Physics \& Astronomy, University of Pennsylvania, 209 South 33rd Street, Philadelphia, PA 19104, USA}

\author[0000-0002-1301-3893]{Yi Mao}
\affiliation{Department of Astronomy, Tsinghua University, Beijing 100084, China}

\begin{abstract}
Intensity mapping of the H{\small I} 21~cm line and the CO 2.61~mm line from the epoch of reionization has emerged as powerful, complementary, probes of the high-redshift Universe. 
However, both maps and their cross-correlation are dominated by foregrounds. 
We propose a new analysis by which the signal is unbiased by foregrounds, i.e.\ it can be measured without foreground mitigation. 
We construct the antisymmetric part of two-point cross-correlation between intensity maps of the H{\small I} 21~cm line and the CO 2.61~mm line, arising because the statistical fluctuations of two fields have different evolution in time. 
We show that the sign of this new signal can distinguish model-independently whether inside-out reionization happens during some interval of time. 
More importantly, within the framework of the excursion set model of reionization, we demonstrate that the slope of the dipole of H{\small I}-CO cross-power spectrum at large scales is linear to the rate of change of global neutral fraction of hydrogen in a manner independent of reionization parameters, until the slope levels out near the end of reionization, but this trend might possibly depend on the framework of reionization modelling. The H{\small I}-CO dipole may be a smoking-gun probe for the speed of reionization, or ``standard speedometer'' for cosmic reionization. Observations of this new signal will unveil the global reionization history from the midpoint to near the completion of reionization. 

\end{abstract}



\section{Introduction} 

Intensity mapping of the 21~cm hyperfine transition line of atomic hydrogen is currently considered to be one of the most promising probes of the epoch of reionization (EOR) \citep{1990MNRAS.247..510S,1997ApJ...475..429M}. Upcoming large radio interferometer arrays promise to detect the 21~cm power spectrum from the EOR for the first time, and will attempt to obtain the tomographic 21~cm imaging (see, e.g.\ \citealt{furlanetto2006cosmology}). As a complementary probe, intensity mapping of the 2.61~mm ($J = 1\to0$) spectral line of the ${}^{12}{\rm CO}$ (carbon monoxide)\citep{2008A&A...489..489R,2011ApJ...730L..30C,gong2011probing,lidz2011intensity,2018MNRAS.473..271V} can probe the gas mass of star-forming regions during the formation of first galaxies. 

\begin{figure*}
\centering
\includegraphics[width=0.65\columnwidth]{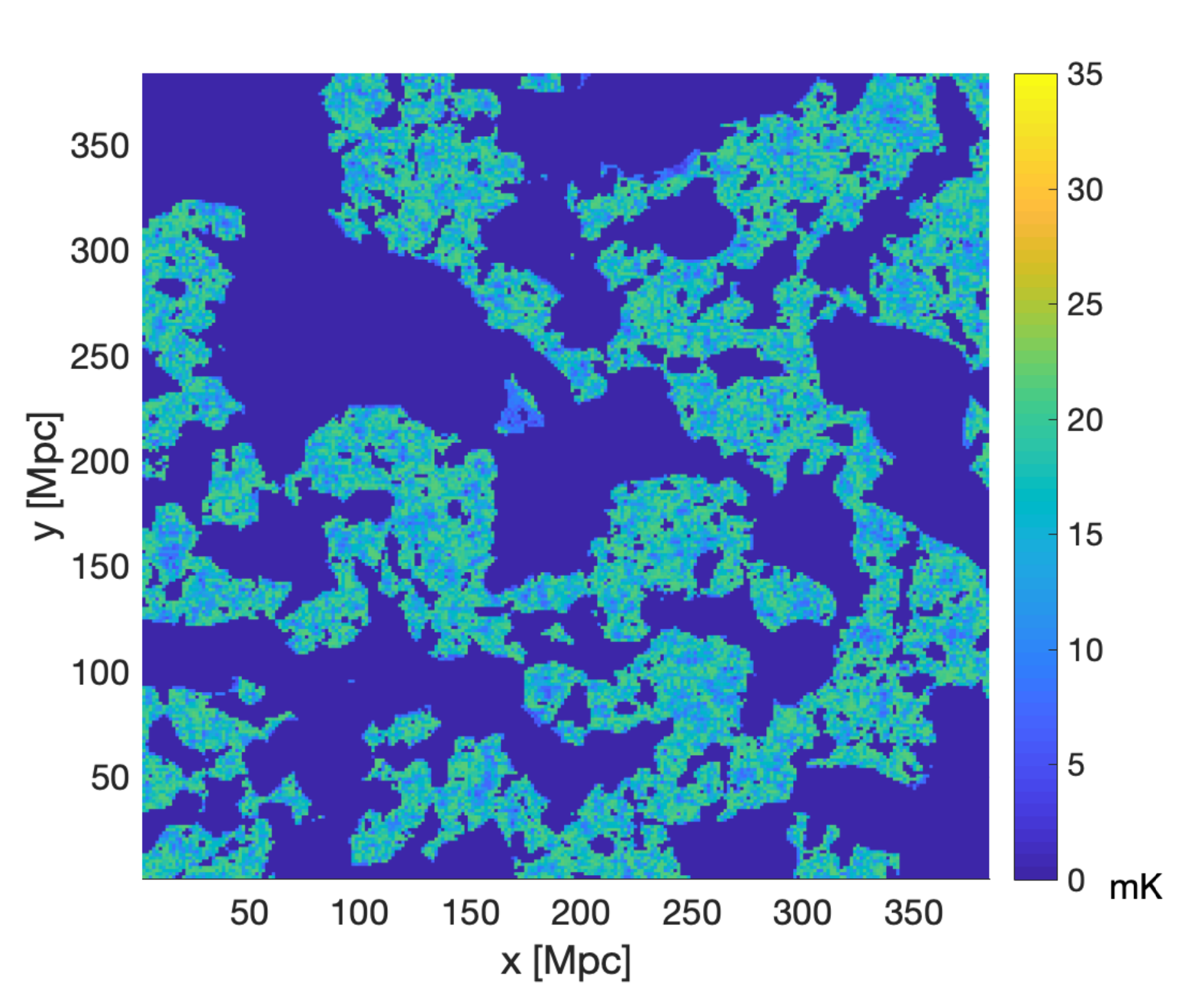} 
\includegraphics[width=0.65\columnwidth]{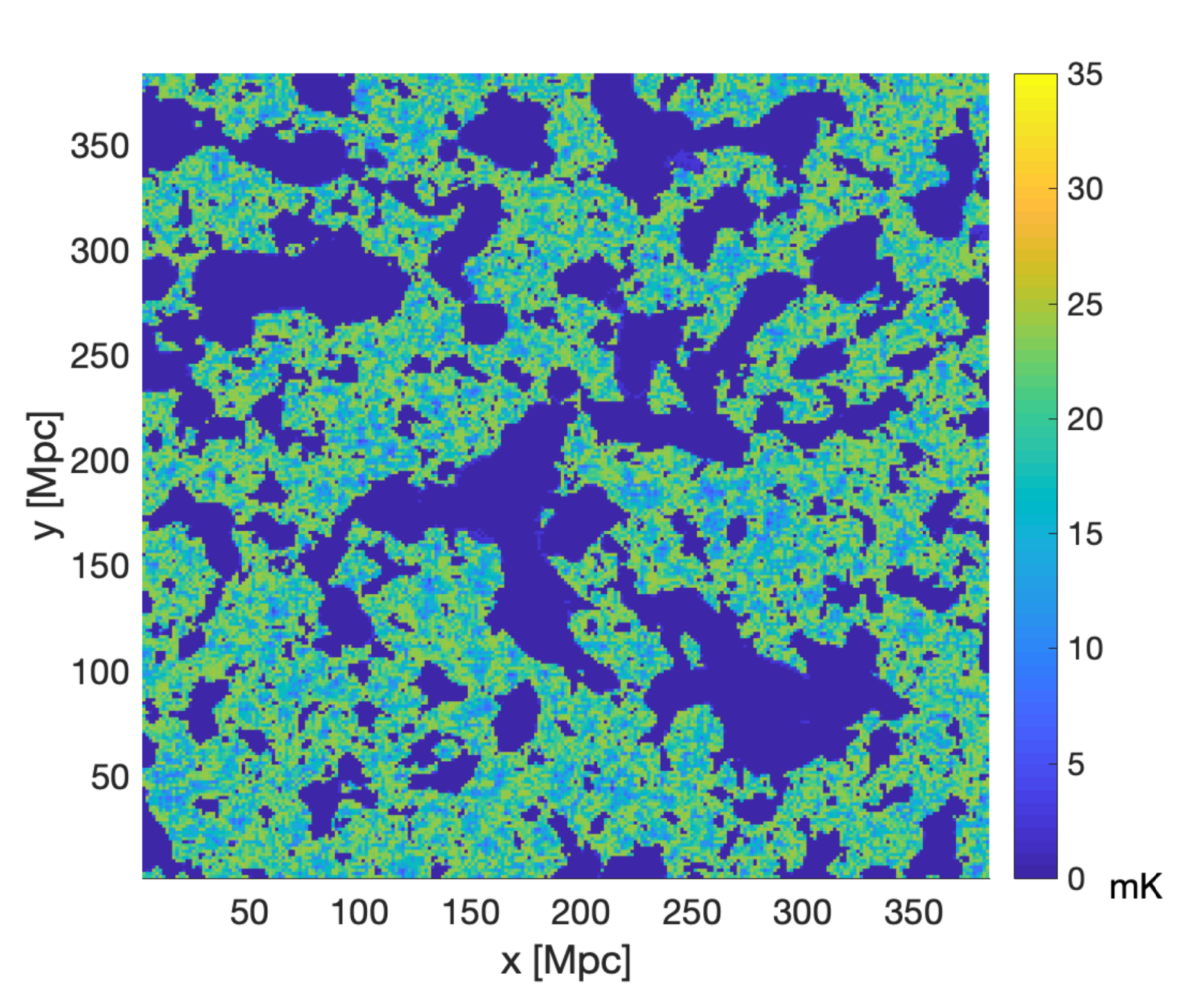}
\includegraphics[width=0.65\columnwidth]{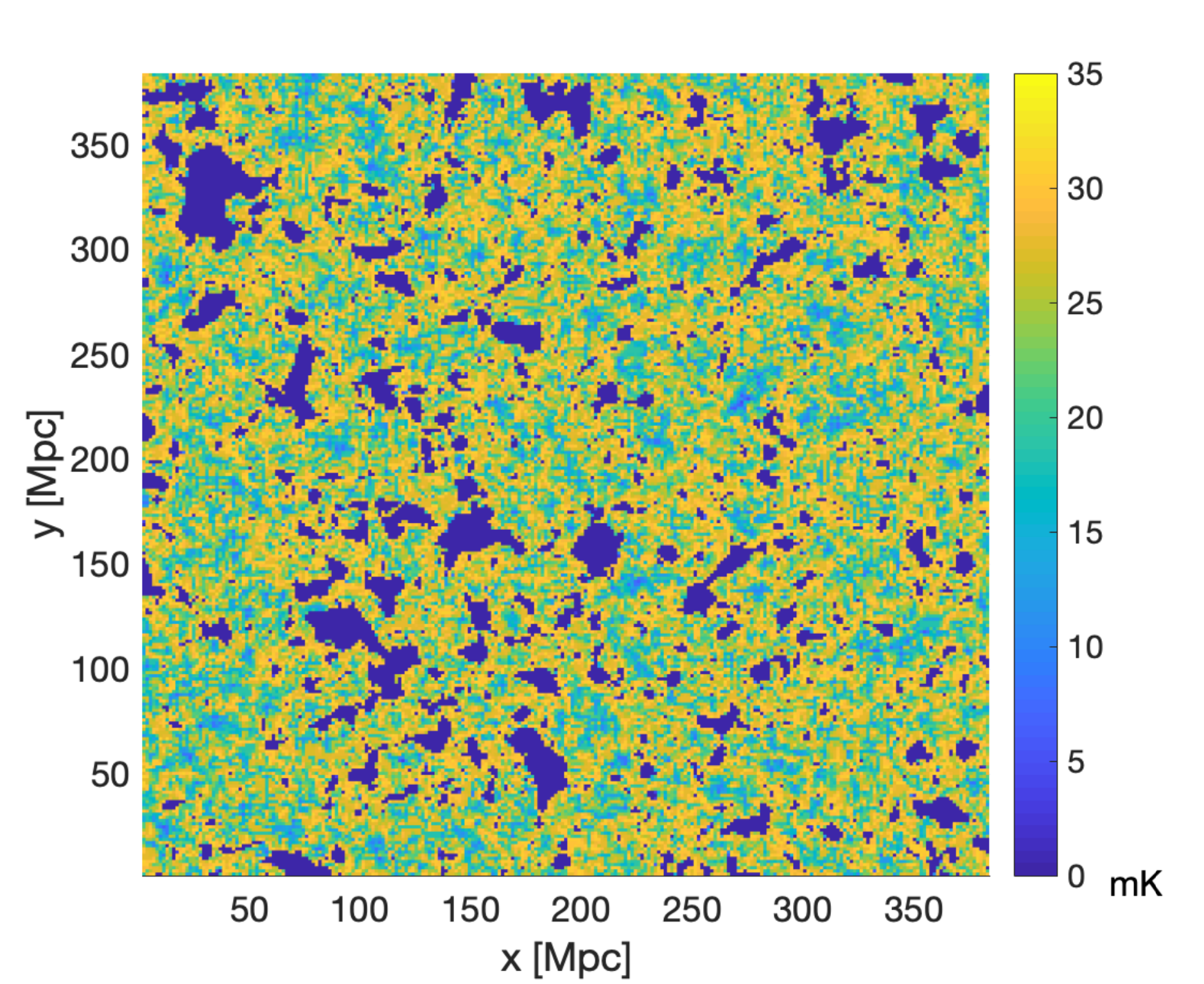}

\includegraphics[width=0.65\columnwidth]{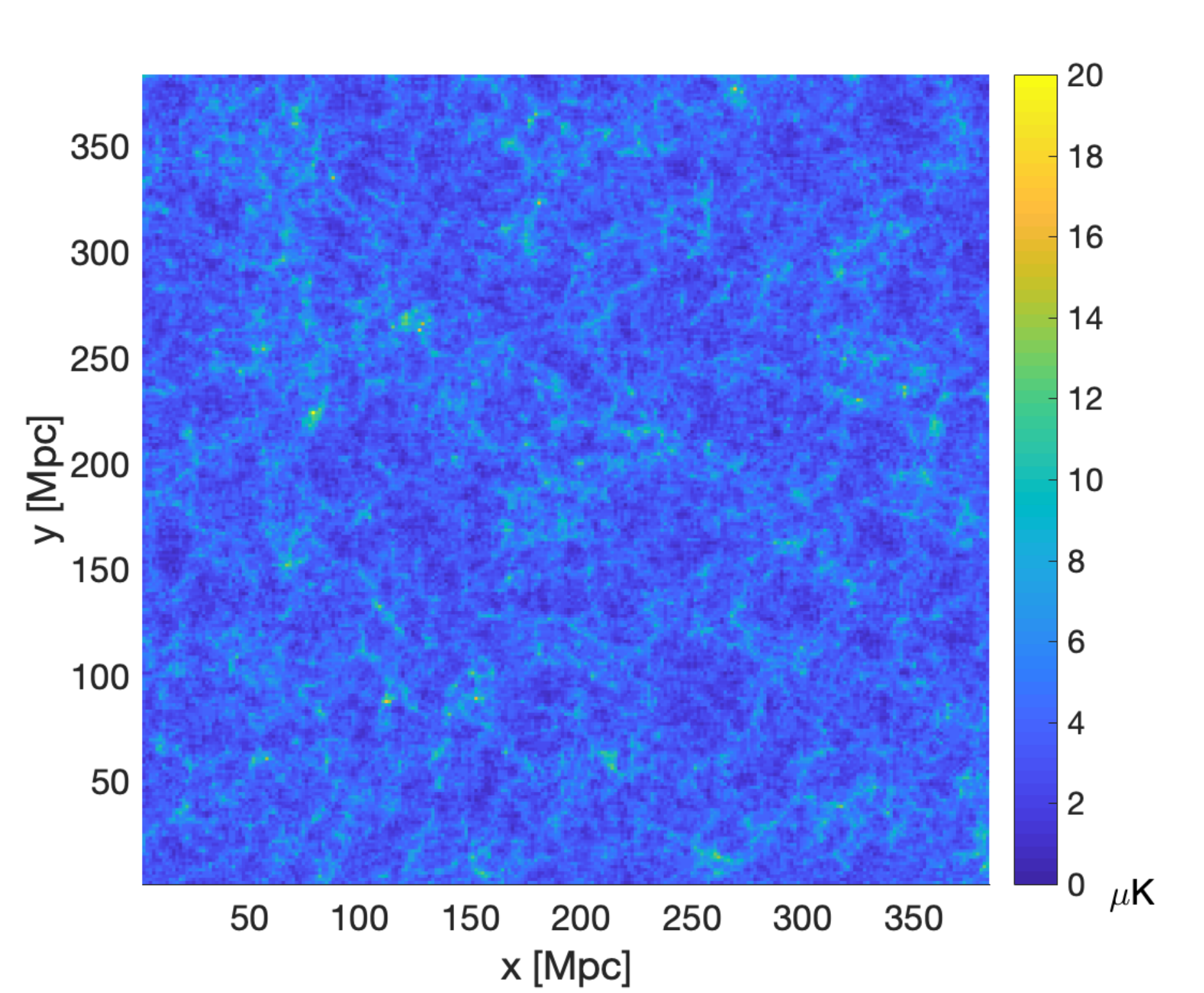} 
\includegraphics[width=0.65\columnwidth]{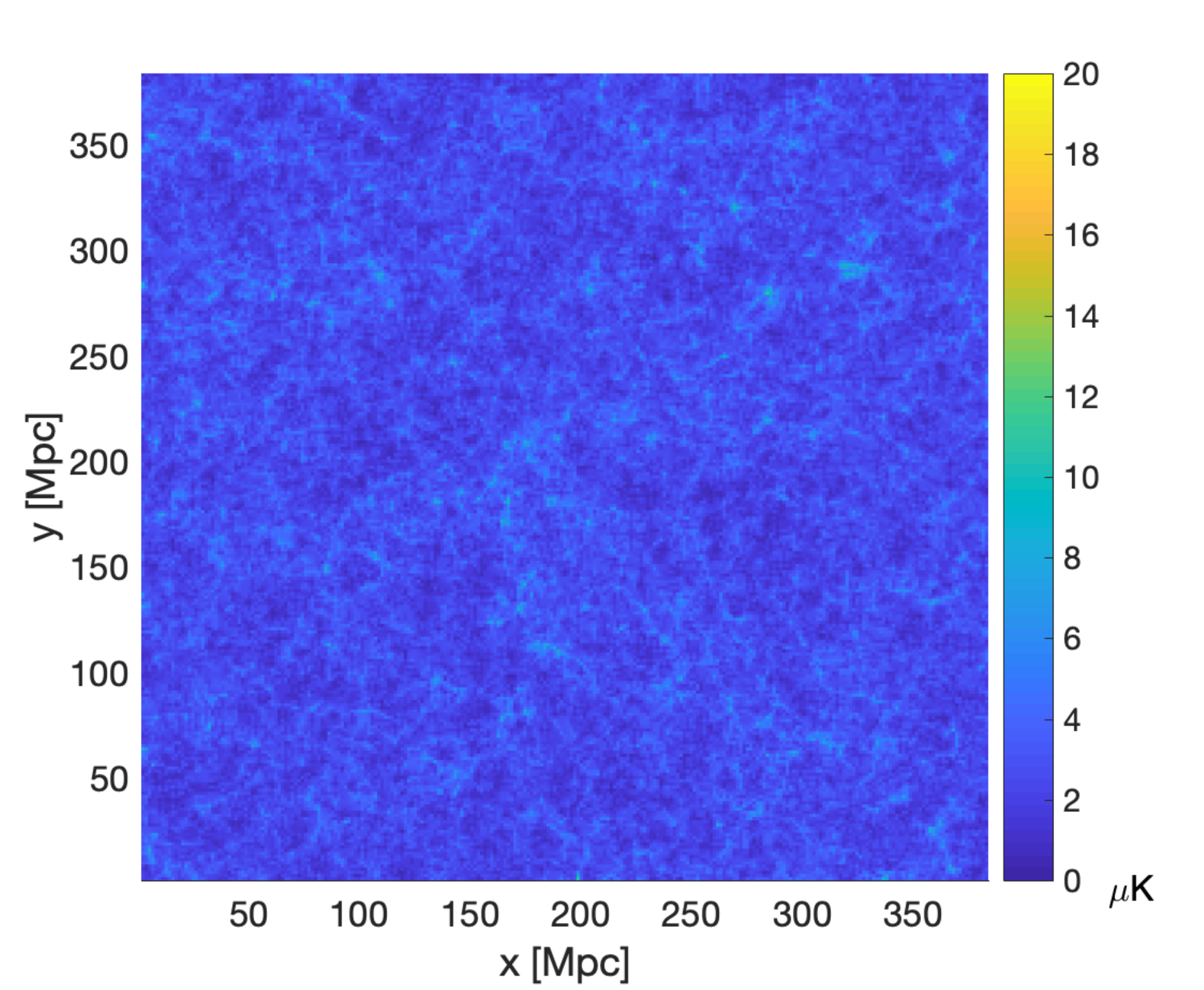}
\includegraphics[width=0.65\columnwidth]{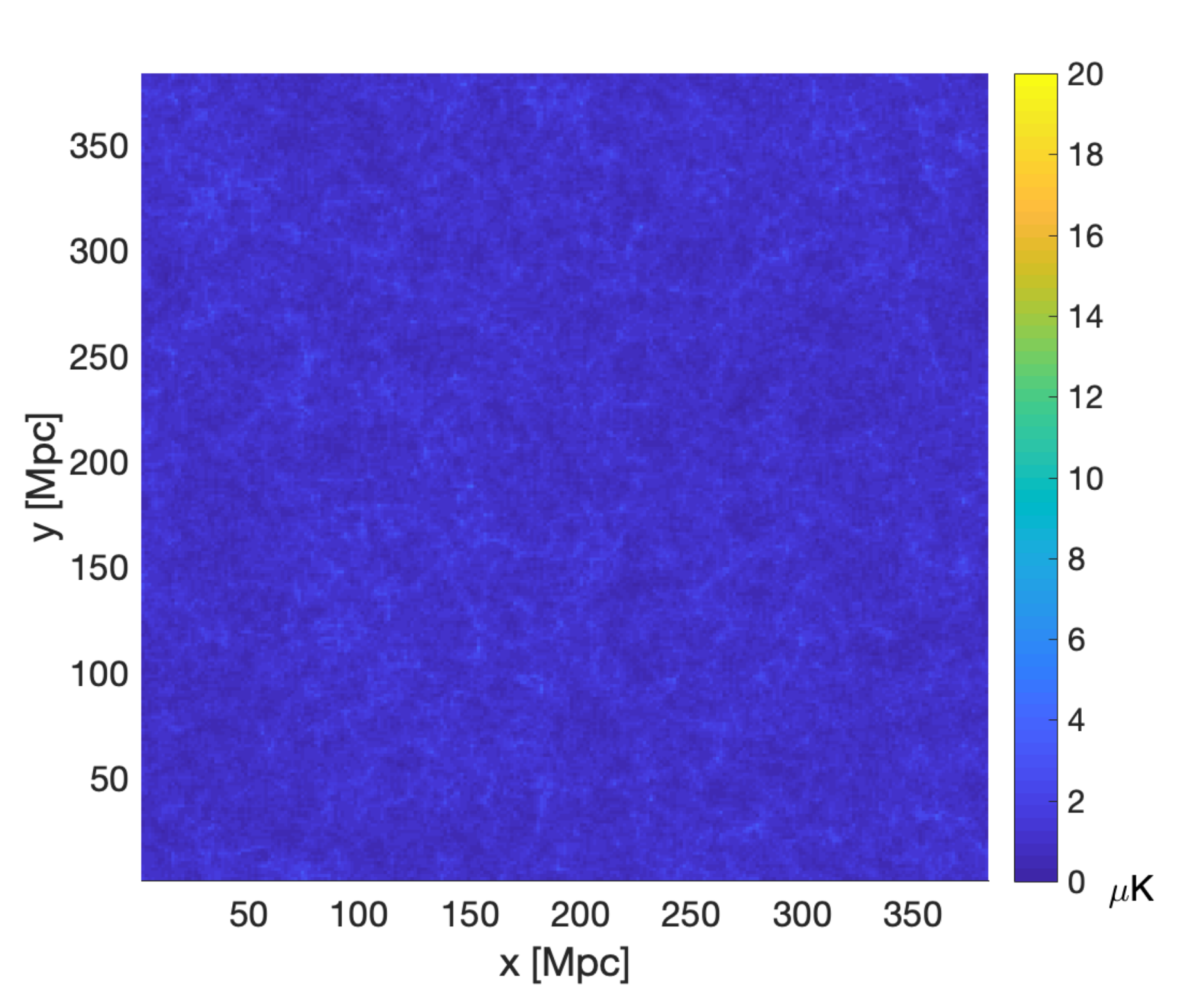}
\caption{Evolution of brightness temperature maps. We show the 21~cm (top) and CO(1-0) maps (bottom) in a slice of simulated lightcone box perpendicular to the LOS in a region of $384$ comoving ${\rm Mpc}$ on each side inside the simulation volume, (from left to right) at redshift $z=7.77$, 8.48, and 9.96 (corresponding to global neutral fraction $\bar{x}_{\rm HI} = 0.25$, $0.50$, and $0.75$ in our fiducial model, i.e.\ with time flow from right to left), respectively.}
\label{Slices}
\end{figure*}

However, intensity maps of both 21~cm line and CO(1-0) line are dominated by foregrounds which are typically orders of magnitude brighter than the signals from the EOR. At low redshifts, two-point cross-correlation of two fields, e.g.\ the H{\small I}-galaxy survey cross-correlation \citep{10.1093/mnras/sty346,10.1046/j.1365-8711.2001.04902.x,Masui_2013,2020MNRAS.495.3935P} or the CO-galaxy survey cross-correlation \citep{2019ApJ...872..186C}, can be robust against foregrounds because the foreground contamination of each field, if any, is usually caused by different sources. In this case, a one-point cross-correlation technique \citep{PhysRevLett.123.231105} is also proposed to be unbiased by foregrounds. Unfortunately, that is not the case at high redshifts for the EOR: the same set of Galactic and extragalactic radio sources contribute to both 21~cm and CO foregrounds at the EOR, so both foregrouds depend on the observed frequency with exactly the same power law, even though their strengths are very different. Cross-correlation between the 21~cm and CO(1-0) line intensity maps \citep{2010JCAP...11..016V,gong2011probing,lidz2011intensity}, consequently, will be affected significantly by foregrounds. Thus sophisticated foreground removal or avoidance techniques (see \citealt{2020PASP..132f2001L} and references therein) must be implemented in order to measure the auto-correlation as well as cross-correlation signals.

In a statistically isotropic universe, the two-point cross-correlation function between any pair of fields $\delta_i$ and $\delta_j$, defined as 
$\xi_{ij}(\textbf{x})= \left<\delta_i(\textbf{x}')\delta_j(\textbf{x}'+\textbf{x}) \right>$,  
is symmetric under the exchange of the line-of-sight (LOS) coordinates or equivalently the order of two fields. 
However, evolution effect \citep{2014PhRvD..89h3535B} can intrinsically break the parity symmetry along the LOS and cause the asymmetry of the cross-correlation in the large-scale structure. This effect is easy to understand: if the field $\delta_i$ evolves more rapidly than the field $\delta_j$, then whether $\delta_i$ is in front of, or behind, $\delta_j$ would result in different cross-correlation for the same physical separation. In principle, larger asymmetry in cross-correlation is expected between two fields with more distinctive evolution in cosmic time. 

Cross-correlation between the H{\small I} 21~cm line and CO(1-0) line intensity mapping from the EOR is indeed asymmetric due to evolution effect. During the EOR, ultraviolet and X-ray photons emitted from the first luminous objects ionize hydrogen atoms first in the surrounding intergalactic medium and form bubbles of ionized hydrogen regions, and eventually these ionized bubbles fill the whole Universe by $z\simeq 6$ \citep{2006ARA&A..44..415F}. 
We illustrate the evolution of the 21~cm and CO line intensity maps with a slice of lightcone box at three representative stages of reionization in Fig.~\ref{Slices}. 
While the CO(1-0) line intensity maps mostly trace the cosmic density distributions that gradually form the filamentary structures, the H{\small I} 21~cm line intensity maps reflect the distributions of neutral hydrogen regions, showing the patchy patterns that rapidly percolate towards the end of reionization. This comparison clearly shows that the progressing of cosmic reionization is faster than the evolution of density fluctuations during the EOR. As such, we expect that the H{\small I}-CO cross-correlation is strongly asymmetric. 
The antisymmetric component of the cross-correlation, 
$\xi_{ij}^{\rm A}(\textbf{x})\equiv\frac{1}{2}\left[\xi_{ij}(\textbf{x})-\xi_{ji}(\textbf{x})\right]$, or equivalently $\xi_{ij}^{\rm A}(\textbf{x})=\frac{1}{2}\left[\xi_{ij}(\textbf{x})-\xi_{ij}(-\textbf{x})\right]$, contains independent statistical information \citep{PhysRevD.93.023507,2017PhRvD..95d3530H} complementary to the symmetric component that the term ``cross-correlation'' was usually implicitly referred to in the literature. 

We propose to extract the antisymmetric cross-correlation between the H{\small I} 21~cm line and the CO 2.61~mm line intensity maps from the EOR as a new probe of cosmic reionization. 
This paper is devoted to explore the characteristics of this new statistical observable, e.g.\ its amplitude and dependence on the wavenumber and the redshift, its dependence on astrophysical parameters, using semi-numerical simulations, and provide our physical interpretation. More importantly, we will investigate the effect of foreground contamination explicitly. 

This paper is organized as follows. In Section~\ref{sec:method}, we describe our reionization simulations, the modelling of the 21~cm and CO line intensity maps, and the methodology of extracting the antisymmetric cross-correlation signal. 
We present the results in Section~\ref{sec:results}, and discuss various dependence and effects of this new signal in Section~\ref{sec:discussions}. In Section~\ref{sec:conclusions} we make concluding remarks. Some derivations and technical details are presented in the Appendix A, B and C.

\section{Methodology}
\label{sec:method}

\subsection{Simulations}
\begin{figure}
\centering
\includegraphics[width=0.9\columnwidth]{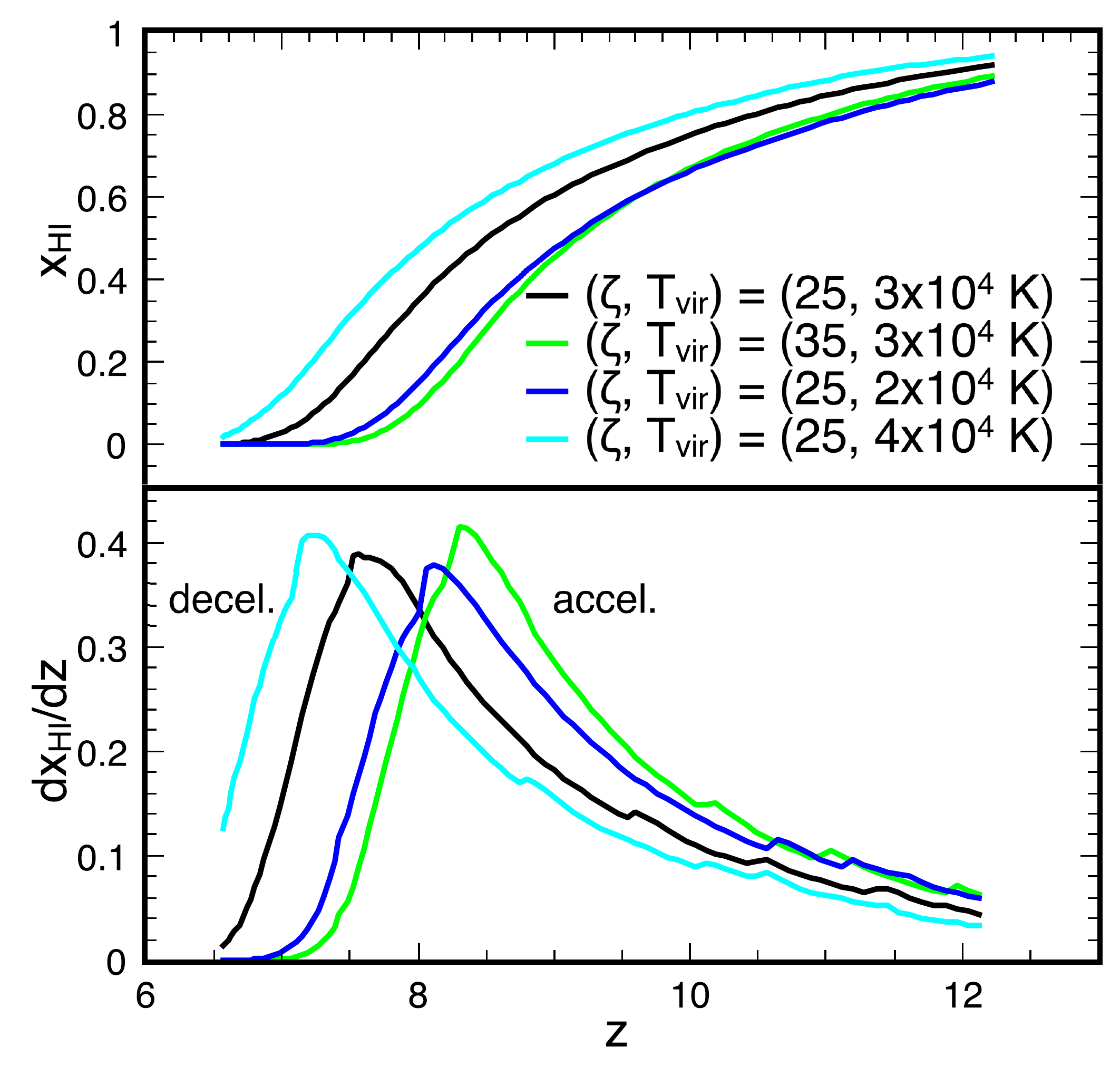}
\caption{Global reionization history for different reionization models as marked in the legend: (top) mean neutral fraction $\bar{x}_{\rm HI}$ vs redshift $z$, and (bottom) its redshift derivative $d\bar{x}_{\rm HI}/dz$ vs $z$. Generically, reionization has two stages, the acceleration (``accel.'') and deceleration (``decel.'') stages.}
\label{AstrophysParameters_globalhistory}
\end{figure}

We perform semi-numerical simulations of reionization with the publicly available code {\tt 21cmFAST}\footnote{\url{https://github.com/21cmfast/21cmFAST}} \citep{Mesinger201121cmfast}.  This code quickly generates the fields of density, velocity, ionized fraction, spin temperature and 21~cm brightness temperature on a grid. It is based on the semi-numerical treatment of cosmic reionization with the excursion-set approach \citep{2004ApJ...613....1F} to identify ionized regions. Specifically, cells inside a spherical region are identified as ionized, if the number of ionizing photons in that region is larger than that of neutral hydrogen atoms. 
Our simulations were performed on a cubic box of 768 comoving ${\rm Mpc}$ on each side, with $512^3$ grid cells. Our EOR model is parametrized with three parameters: $\zeta$ (the ionizing efficiency), $T_{{\rm vir}}$ (the minimum virial temperature of halos that host ionizing sources), and $R_{\rm mfp}$ (the maximum mean free path of ionizing photons).  
For the purpose of illustration, we choose a reference case with the parameter values $\zeta = 25$, $T_{\mathrm{vir}} = 3 \times 10^{4}\,{\rm K}$, $R_{\mathrm {mfp}} = 50 \,{\rm Mpc}$. This yields a global reionization history in which the reionization starts at $z\simeq 16$ and ends at $z\simeq 6.5$, with the cosmic microwave background (CMB) Thomson scattering optical depth $\tau =  0.068$, consistent with current observational constraints on the reionization history \citep{2017MNRAS.465.4838G}. 
To shed light on the nature of the H{\small I}-CO dipole, we further consider a wider range of astrophysical parameters in the EOR model as listed in the legend of Fig.~\ref{AstrophysParameters_globalhistory}. In all cases, $R_{\mathrm {mfp}} = 50 \,{\rm Mpc}$ is fixed because the results depend on this parameter very weakly (see Appendix \ref{app:Rmfp} for the detail). 
Fig.~\ref{AstrophysParameters_globalhistory} shows that these models result in very different reionization histories as well as different rates of change of mean neutral fraction, $d\bar{x}_{\rm HI}/dz$, or the ``speed'' of reionization. 

In this paper, we adopt the standard $\Lambda \mathrm{CDM}$ cosmology with fixed values of cosmological parameters based on the Planck 2016 results \citep{2016A&A...594A..13P}, $\left(h, \Omega_m, \Omega_b, \Omega_{\Lambda}, \sigma_{8}, n_{\mathrm{s}}\right)=(0.678,0.306,0.0492,0.694,0.82,0.968)$. 

\begin{figure}
\centering
\includegraphics[width=0.8\columnwidth]{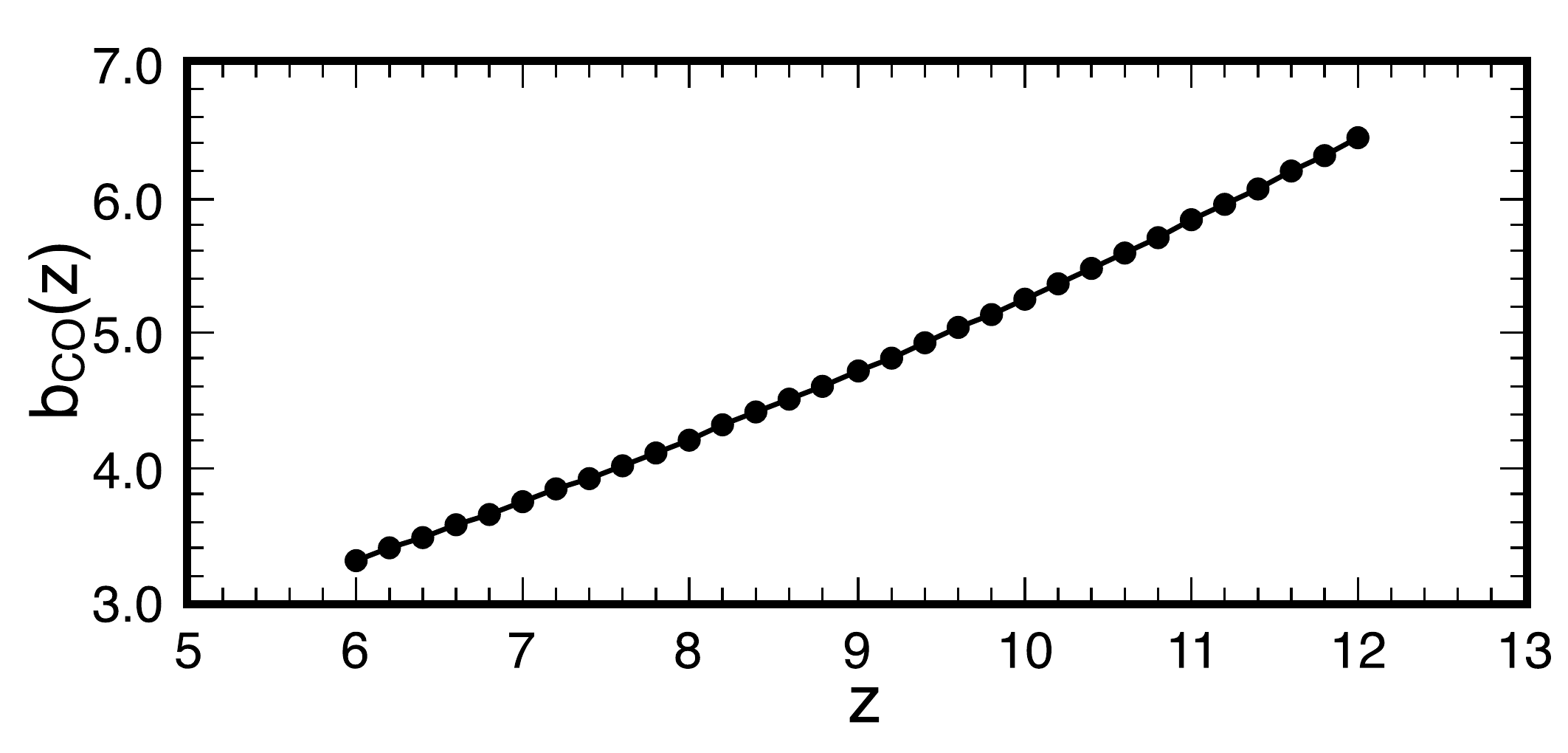}
\caption{Evolution of the CO bias. We show the CO bias $b_{\rm CO}$ vs redshift $z$ in our fiducial EOR model.}
\label{bCO}
\end{figure}

\subsection{Modelling the brightness temperature} 
The 21~cm brightness temperature at position ${\bf x}$ relative to the CMB temperature can be written \citep{furlanetto2006cosmology} as 
\begin{equation}
T_{21}(\textbf{x},z)=\tilde{T}_{21}(z)\,x_{\rm HI}(\textbf{x})\,\left[1+\delta(\textbf{x})\right]\,(1-\frac{T_{\rm CMB}}{T_S})\,,\label{eqn:21cm}
\end{equation}
where $\tilde{T}_{21}(z) = 27\sqrt{[(1+z)/10](0.15/\Omega_m h^2)}(\Omega_b h^2/0.023)$ in units of mK. 
Here, $x_{\rm HI}({\bf x})$ is the neutral fraction, and $\delta({\bf x})$ is the matter overdensity, at position ${\bf x}$. We assume the baryon distribution traces the cold
dark matter on large scales, so $\delta_{\rho_{\rm H}} = \delta$. 
In this paper, we focus on the limit where spin temperature $T_S \gg T_{\rm CMB}$, valid soon after reionization begins. As such, we can neglect the dependence on spin temperature. Also, for simplicity, we ignore the effect of peculiar velocity, because it only weakly affects the light-cone effect. 

The CO(1-0) line specific intensity can be written \citep{gong2011probing} as 
$I_{\rm CO}(\textbf{x},z)=\bar{I}_{\rm CO}(z)\,\left[1+b_{\rm CO}(z)\,\delta(\textbf{x})\right]$ 
with the mean intensity $\bar{I}_{\rm CO}(z)$ and the bias $b_{\rm CO}(z)$ at redshift $z$. 
The equivalent brightness temperature is computed using the Rayleigh-Jeans Law, 
$T_{\mathrm{CO}}=c^{2} I_{\rm CO} /(2 k_{B} \nu_{\mathrm{obs}}^{2})$. Therefore, it can be written as 
\begin{equation}
   T_{\rm CO}(\textbf{x},z)=\bar{T}_{\rm CO}(z)\,\left[1+b_{\rm CO}(z)\,\delta(\textbf{x})\right]\,. 
   \label{eqn:COtemp}
\end{equation}
Here the observed frequency $\nu_{\mathrm{obs}}=\nu_{\rm CO}/(1+z)$ for gas at redshift $z$ emitting in the CO(1-0) line with the rest-frame $\nu_{\rm CO} = c/\lambda_{\rm CO}=115\,{\rm GHz}$ and $\lambda_{\rm CO}=2.61\,{\rm mm}$.

The mean intensity \citep{gong2011probing,lidz2011intensity} is
$\bar{I}_{\rm CO}(z) = (\lambda_{\rm CO}/4\pi\,H(z)) \int_{M_{\rm CO,min}}^{\infty}\,dM\,(dn/dM)(M,z)\,L_{\rm CO}(M)$, 
where $dn/dM$ is the halo mass function \citep{1999MNRAS.308..119S} and $L_{\rm CO}(M)$ is the CO(1-0) luminosity which is assumed to be a function of the halo mass $M$. We follow the modelling of \cite{lidz2011intensity} in which the CO luminosity is linear to halo mass, 
$	L_{\mathrm{CO}}(M)=2.8 \times 10^{3} L_{\odot}\,(M/10^{8} M_{\odot})$. 
Under this assumption, the mean brightness temperature is  \citep{lidz2011intensity} 
\begin{equation}
\bar{T}_{\rm CO}(z)=59.4\,\mu{\rm K}\,(1+z)^{1/2} f_{\rm coll}(M_{\rm CO, min};z)\,,
\end{equation}
where $f_{\rm coll}(M_{\rm CO, min};z)$ is the mean collapse fraction at $z$ with the lower mass cutoff at $M_{\rm CO, min}$. 
We assume that $M_{\rm CO,min}$, the minimum mass of halos that can host galaxies, is the same as the mass scale of atomic hydrogen cooling. In other words, $M_{\rm CO,min}$ corresponds to $T_{\rm vir}$, the minimum virial temperature of halos that can host ionizing sources. 

The bias $b_{\rm CO}(z)$ describes how well the CO brightness temperature fluctuations trace the matter density fluctuations. It can be modelled  \citep{gong2011probing,lidz2011intensity}  as 
\begin{equation}
b_{\rm CO}(z)=\frac{\int_{M_{\rm CO, min}}^{\infty}\,dM\,\frac{dn}{dM}(M,z)\,L_{\rm CO}(M)\,b(M,z)}
{\int_{M_{\rm CO, min}}^{\infty}\,dM\,\frac{dn}{dM}(M,z)\,L_{\rm CO}(M)} \,,
\nonumber 
\end{equation}
where $b(M,z)$ is the halo bias \citep{1999MNRAS.308..119S}. The linear luminosity assumption can further simplify this expression.  
Instead of evaluation by direct numerical integration, we find an analytic form for the bias $b_{\rm CO}(z)$. We take the form of halo bias,  
$b(M,z) = 1+(\nu-1)/\delta_c(z)$, where $\nu = \delta_c^2/\sigma^2(M)$, by setting $a=1$ and $p=0$ in Eq.~(12) of \citet{1999MNRAS.308..119S}. In this case, the mass function is the Press-Schechter form. 
Thus we can obtain an analytical form, 
\begin{eqnarray}
   b_{\rm CO}(z) &=& 1-\frac{1}{\delta_c(z)}+\frac{2}{\sqrt{\pi}\delta_c (z) f_{\rm coll}(M_{\rm CO,min};z)}\nonumber \\
   & & \times\, \Gamma_{\rm inc}\left(1.5,\frac{\delta^2_c(z)}{2\sigma^2_{\rm min}}\right) \,,
   \label{eqn:biasCO-appen}
\end{eqnarray}
where the incomplete Gamma function $\Gamma_{\rm inc}(a,x) \equiv \int_{x}^{\infty} t^{a-1} e^{-t} d t $, and $\sigma_{\rm min} \equiv \sigma({M_{\rm CO,min}})$. 
In Fig.~\ref{bCO}, we find that the CO bias decreases with time, a trend also found in \cite{gong2011probing}. 

\subsection{Dipole}
We use the dipole of the H{\small I}-CO cross-power spectrum as the antisymmetry estimator. The antisymmetric component of the cross-correlation between two fields $\delta_i$ and $\delta_j$ is Fourier dual to the imaginary part of the cross-power spectrum, 
\begin{equation}
\frac{1}{2i} \left[\left<\widetilde{\delta_i}(\textbf{k}_1)\widetilde{\delta_j}^\ast(\textbf{k}_2)\right>-{\rm c.c.}\right] \equiv (2\pi)^3\delta_{\rm D}^{(3)}(\textbf{k}_1-\textbf{k}_2)P_{ij}^{\rm I}(\textbf{k}_1)\,.
\end{equation} 
Here ``c.c.'' stands for the complex conjugate of the first term, and $\widetilde{\delta}(\textbf{k})$ is the Fourier dual to the field $\delta(\textbf{x})$. Since $P_{ij}^{\rm I}(\textbf{k}) = - P_{ji}^{\rm I}(\textbf{k})$, $P_{ij}^{\rm I}(\textbf{k})$ is called the ``antisymmetric cross-power spectrum''. 
Note that $P_{ij}^{\rm I}(-\textbf{k}) = - P_{ij}^{\rm I}(\textbf{k})$, i.e. $P_{ij}^{\rm I}(\textbf{k})$ is also antisymmetric in ${\bf k}$-space, so its averaging over a spherical $k$-shell (i.e.\ monopole) is zero, and only odd moments are nonzero. The leading order in terms of the expansion in spherical harmonics is the dipole. Thus we neglect higher order terms and assume a simple template for extracting the dipole $P_{ij}^{\rm I}(\textbf{k}) = P_{ij}^{\rm A}(k)\,Y_{10}(\theta)$. 
At a given spherical $k$-shell, we first average $P_{ij}^{\rm I}(\textbf{k})$ over polar angles in a ring with the same azimuthal angle $\theta$ with respect to the LOS ($z$-axis), and obtain the average, $P_{ij}^{\rm I}(k,\theta)$. It is straightforward to derive the variance of $P_{ij}^{\rm I}(k,\theta)$ (see Appendix~\ref{app:var} for the derivation), 
\begin{eqnarray}
    \sigma^2_{P_{ij}^{\rm I}}(k,\theta) &=& \frac{1}{N(k,\theta)}\left[P_{i}(k)\,P_{j}(k)-\left(P_{ij}^{\rm S}(k)\right)^2\right. \nonumber\\
    & & \left.+\left(P_{ij}^{\rm I}(k,\theta)\right)^2\right]\,,\label{eq:var}
\end{eqnarray}
where $P_{i}(k)$ and $P_{j}(k)$ are the auto-power spectrum of the fields $\delta_i$ and $\delta_j$, respectively, and $P_{ij}^{\rm S}(k)$ is their symmetric (i.e.\ real part of) cross-power spectrum. The auto-power and the symmetric cross-power spectrum only depend on the magnitude of wavenumber $k$ because the lightcone effect for each separate statistics is not important \citep{2012MNRAS.424.1877D}. Here $N(k,\theta)$ is the number of cells in the ring of $(k,\theta)$ in the upper hemisphere. We test that $\sigma^2_{P_{ij}^{\rm I}}(k,\theta)$ is always real and positive. Next, the dipole $P_{ij}^{\rm A}(k)$ and its variance can be estimated with $\chi^2$-fitting over all measures at different $\theta$ in the spherical $k$-shell. The estimator of the dipole is 
\begin{equation}
P_{ij}^{\rm A}(k) = \sigma^2_{P_{ij}^{\rm A}}(k)\,\sum_{\theta}\frac{P_{ij}^{\rm I}(k,\theta)\,Y_{10}(\theta)}{\sigma_{P_{ij}^{\rm I}}^2(k,\theta)}\,,
\end{equation}
where the variance of the dipole estimation is 
\begin{equation}
\sigma^2_{P_{ij}^{\rm A}}(k) = \left[\sum_{\theta}\frac{Y_{10}^2(\theta)}{\sigma_{P_{ij}^{\rm I}}^2(k,\theta)}\right]^{-1}\,.
\end{equation}

\subsection{Mock signal}
We generate realizations of the 21~cm and CO brightness temperature fields using the density and ionized fraction data from semi-numerical simulations with the code {\tt 21cmFAST}. We then interpolate the snapshots at different time to construct the lightcone data cube along the LOS. To reduce the interpolation error caused by insufficient sampling of snapshots, we output the simulation results at 100 different redshifts from $z= 12.22$ to 6.56 (corresponding to $\bar{x}_{\rm HI} = 0.95$ to 0.01 in our fiducial model) in such a manner that these redshifts correspond to the comoving LOS distances with equal separation of 13.7 comoving Mpc. 
To avoid the impact of periodic boundary condition, we only use the inner cubic region of $384$ comoving ${\rm Mpc}$ on each side away from the boundary of the simulation box (see Appendix~\ref{app:conv} for convergence tests and justification of choosing this size). 
To mimic the observations from radio interferometers, we subtract from the lightcone field the mean of the 2D slice for each 2D slice perpendicular to the LOS, because radio interferometers cannot measure the mode with ${\bf k}_\perp =0 $. This forms the mock lightcone fields of $\delta T_{21}(\textbf{x})$ and $\delta T_{\rm CO}(\textbf{x})$ in a cubic box with its location marked by its central redshift $z$. The Fourier transforms of the two fields are used to compute the cross-power spectrum and then its dipole, $P_{\rm H{\small I}-CO}^{\rm A}(k)$, using the aforementioned prescription. 
Note that the antisymmetric cross-power spectrum flips its sign if the order of cross-correlation is swapped, so we fix our convention that the order of cross-correlation is H{\small I}-CO throughout the paper, and thus the subscript ``H{\small I}-CO'' in the notation of the dipole $P_{\rm H{\small I}-CO}^{\rm A}(k)$ is dropped for the rest of this paper. Finally, we vary the initial condition and generate 100 different realizations, in order to lower the cosmic variance of dipole from simulations. 

\begin{figure}
\centering
\includegraphics[width=0.9\columnwidth]{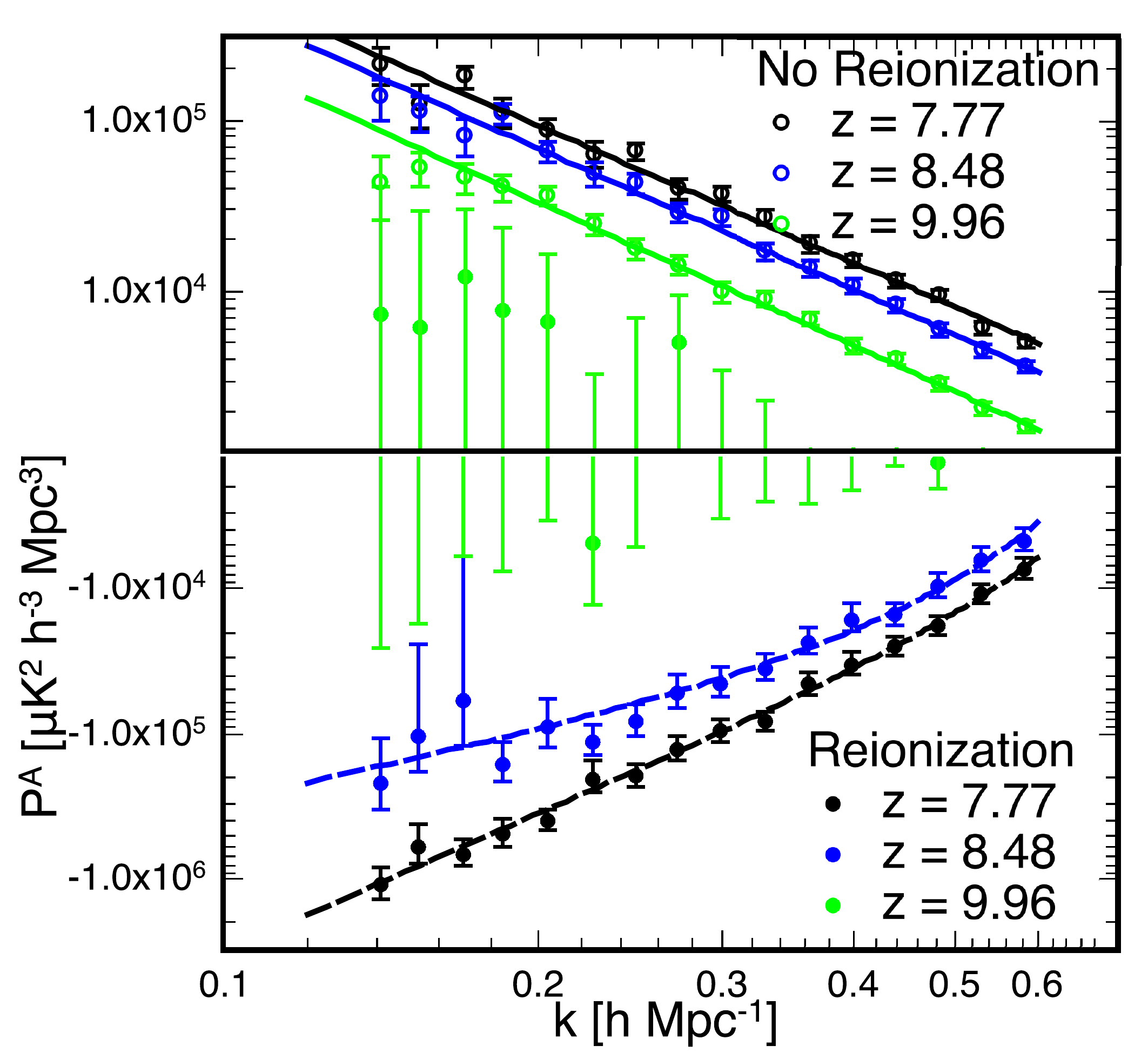} 
\caption{The dipole of the H{\small I}-CO cross-power spectrum, $P^{\rm A}$, vs wavenumber $k$ in our fiducial reionization model, when the center of lightcone box is at $z=9.96$ (green solid dots), $8.48$ (blue solid dots), and $7.77$ (black solid dots), respectively. We fit the dipole to a modified power law (dashed lines). For illustrative purpose, we also show the results in the hypothetical no-reionization scenario (open dots) and fit the dipole to a power law (solid lines) at each central redshift. The error bars are 1$\sigma$ standard deviation for cosmic variance corresponding to the simulation volume of 100 realizations.}
\label{MainResults}
\end{figure}

\begin{figure}
\centering
\includegraphics[width=0.7\columnwidth]{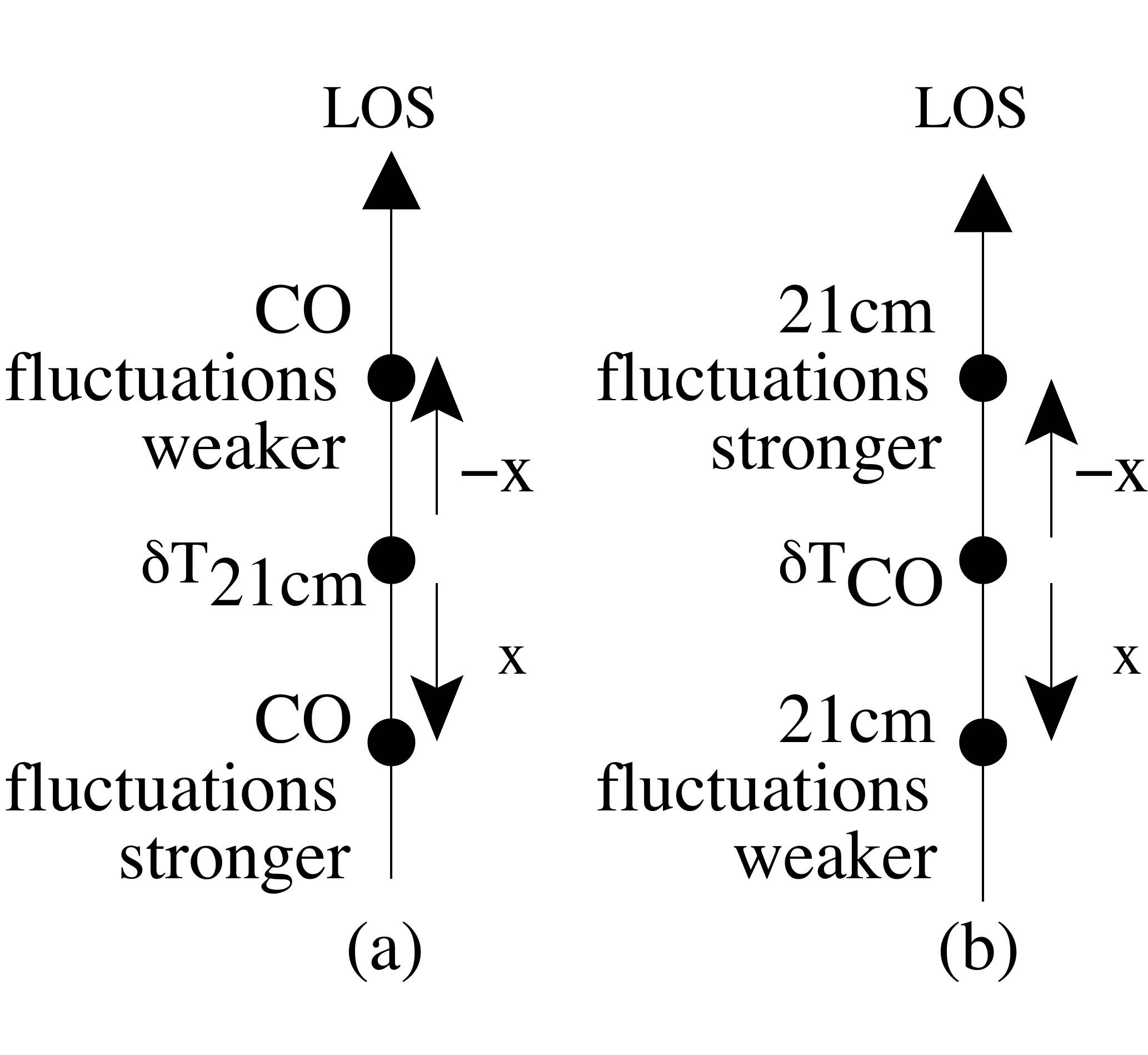} 
\caption{Illustration of the evolution effect for antisymmetric cross-correlation, in a universe (a) with no reionization, and (b) with inside-out reionization. }
\label{schematic}
\end{figure}

\begin{figure}
\centering
\includegraphics[width=0.9\columnwidth]{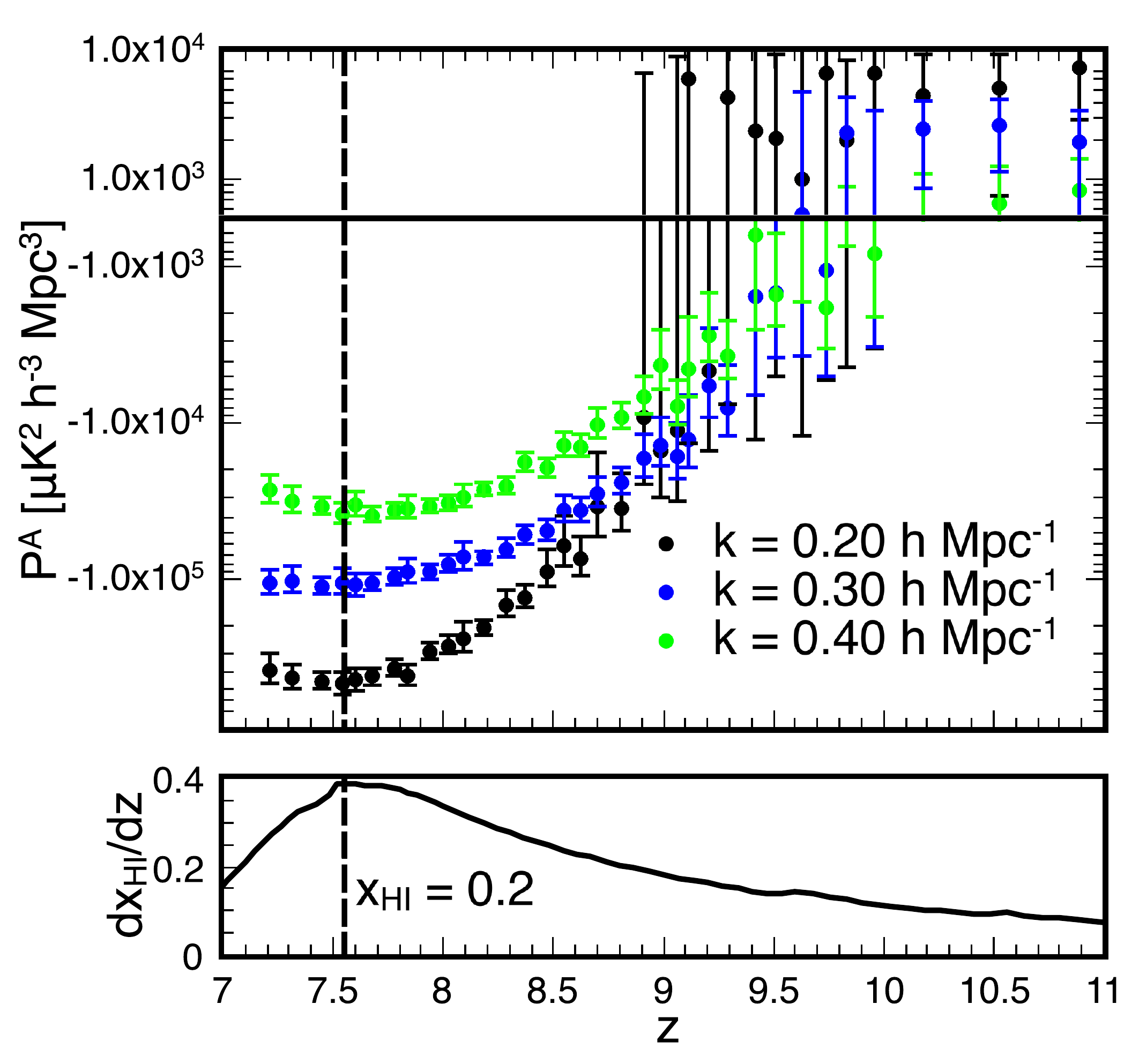}
\caption{(Top) the dipole of the H{\small I}-CO cross-power spectrum, $P^{\rm A}(k,z)$, as a function of redshift $z$ for our fiducial reionization model when the wavenumber is at $k=0.20$  $h\,{\rm Mpc}^{-1}$ (black), 0.30  $h\,{\rm Mpc}^{-1}$ (blue), 0.40 $h\,{\rm Mpc}^{-1}$ (green), respectively. (Bottom) the redshift derivative of mean neutral fraction $d\bar{x}_{\rm HI}/dz$ as a function of redshift $z$ for our fiducial reionization model. Both $P^{\rm A}(z)$ and $d\bar{x}_{\rm HI}/dz$ reach their extrema at the same redshift $z = 7.55$, as marked by the vertical dashed line, corresponding to $\bar{x}_{\rm HI} = 0.2$ in our fiducial model.}
\label{Evolution}
\end{figure}

\section{Results}
\label{sec:results}

In Fig.~\ref{MainResults} we plot the dipole of the H{\small I}-CO cross-power spectrum $P^{\rm A}(k)$. 
If, hypothetically, the universe was not reionized, i.e.\ $x_{\rm HI}=1$ everywhere all the time, we find that the H{\small I}-CO dipole is generically positive, and it fits to a power law, $P^{\rm A}(k)= A_{\rm NR} (k/k_\ast)^{-n_{\rm NR}}$ at each redshift of the box center (hereafter ``central redshift''), where we choose $k_\ast = 1\,h\,{\rm Mpc}^{-1}$. 
On the other hand, in the fiducial reionization simulation, while the signal is initially too small to be distinguished from its cosmic variance at high redshifts, we find that the dipole is generically negative, and it fits to a power law at large scales with suppression at small scales, 
\begin{equation}
P^{\rm A}(k)= -A_{\rm R} (k/k_\ast)^{-n_{\rm R}} \exp{\left[-\beta_{\rm R}(k/k_\ast)^{\alpha_{\rm R}} \right]}\,,
\label{eqn:dipole_R}
\end{equation}
at each central redshift. The coefficients $A_{\rm R}$ and $n_{\rm R}$ quantify the power law at large scales, while $\beta_{\rm R}$ and $\alpha_{\rm R}$ quantify the exponential suppression at small scales.

The sign of the antisymmetric cross-correlation can be explained by the evolution effect as illustrated in Fig.~\ref{schematic}. 
In the hypothetical no-reionization scenario, the 21~cm fluctuations are simply equal to the matter density fluctuations which evolve relatively slowly during the EOR. Since the CO bias decreases with time \citep{gong2011probing}, the CO brightness temperature fluctuations decrease with time statistically, too. Therefore, the H{\small I}-CO cross-correlation is stronger in the back end than in the front end, i.e.\ $ \left<\delta_{\rm 21cm}(\textbf{x}')\delta_{\rm CO}(\textbf{x}'+\textbf{x}) \right> >  \left<\delta_{\rm 21cm}(\textbf{x}')\delta_{\rm CO}(\textbf{x}'-\textbf{x}) \right>$. Thus the H{\small I}-CO antisymmetric cross-correlation is positive, or $\xi_{\rm H{\small I}-CO}^{\rm A}(\textbf{x}) > 0$. On the other hand, in the reionization simulations, the universe is reionized with ``inside-out'', i.e.\ overdense regions are reionized earlier on average, so the 21~cm and CO fluctuations are anti-correlated (i.e.\ their cross-correlation is negative) at large scales. Since reionization proceeds much faster than the CO bias evolution, the evolution effect due to reionization dominates over that due to the change of CO bias. Since the 21~cm fluctuations grow with time, the CO-H{\small I} cross-correlation is weaker in the back end than in the front end, i.e.\ $ \left| \left<\delta_{\rm CO}(\textbf{x}')\delta_{\rm 21cm}(\textbf{x}'+\textbf{x}) \right>\right| <  \left|\left<\delta_{\rm CO}(\textbf{x}')\delta_{\rm 21cm}(\textbf{x}'-\textbf{x}) \right> \right|$, so $ \left<\delta_{\rm CO}(\textbf{x}')\delta_{\rm 21cm}(\textbf{x}'+\textbf{x}) \right> >  \left<\delta_{\rm CO}(\textbf{x}')\delta_{\rm 21cm}(\textbf{x}'-\textbf{x}) \right> $, thus $\xi_{\rm CO-H{\small I}}^{\rm A}(\textbf{x}) > 0$. So the H{\small I}-CO antisymmetric cross-correlation is negative, or  $\xi_{\rm H{\small I}-CO}^{\rm A}(\textbf{x}) < 0$. This picture is applied for a general class of inside-out reionization models, so the sign of the H{\small I}-CO antisymmetric cross-correlation can tell whether or not inside-out reionization happens during some interval of time, regardless of the detail of reionization models. 

While Figures~\ref{MainResults} shows that the magnitude of the H{\small I}-CO dipole $P^{\rm A}(k,z)$ increases with time for the central redshift $7.77 \le z \le 9.96$, to further investigate the evolution of the dipolar signal over the entire history of reionization, we plot $P^{\rm A}(k,z)$ as a function of $z$ for fixed wavenumbers in our fiducial reionization model in Figure~\ref{Evolution}. We find that the magnitude of dipole does not always increase monotonically. Instead, it reaches the maximum at the same central redshift $z = 7.55$ for all wavenumbers, and the change from $z\sim 9$ to $z = 7.55$ can be about an order of magnitude larger. Observation of the H{\small I}-CO dipole at this maximum point may have the largest signal-to-noise ratio. 

In search for an interpretation of the maximum, we find an interesting fact that the derivative of mean neutral fraction with respect to the redshift, $d\bar{x}_{\rm HI}/dz$, also reaches the maximum at the same redshift $z = 7.55$ (which corresponds to $\bar{x}_{\rm HI} = 0.2$ in our fiducial model), as shown in Figure~\ref{Evolution}. This coincidence seems to indicate that the dipole is sensitively correlated to the  speed of reionization, $d\bar{x}_{\rm HI}/dz$. We will have more discussions on this correlation in Section~\ref{subsec:model-indep}. 

\section{Discussions}
\label{sec:discussions}

\begin{figure}
\centering
\includegraphics[width=0.9\columnwidth]{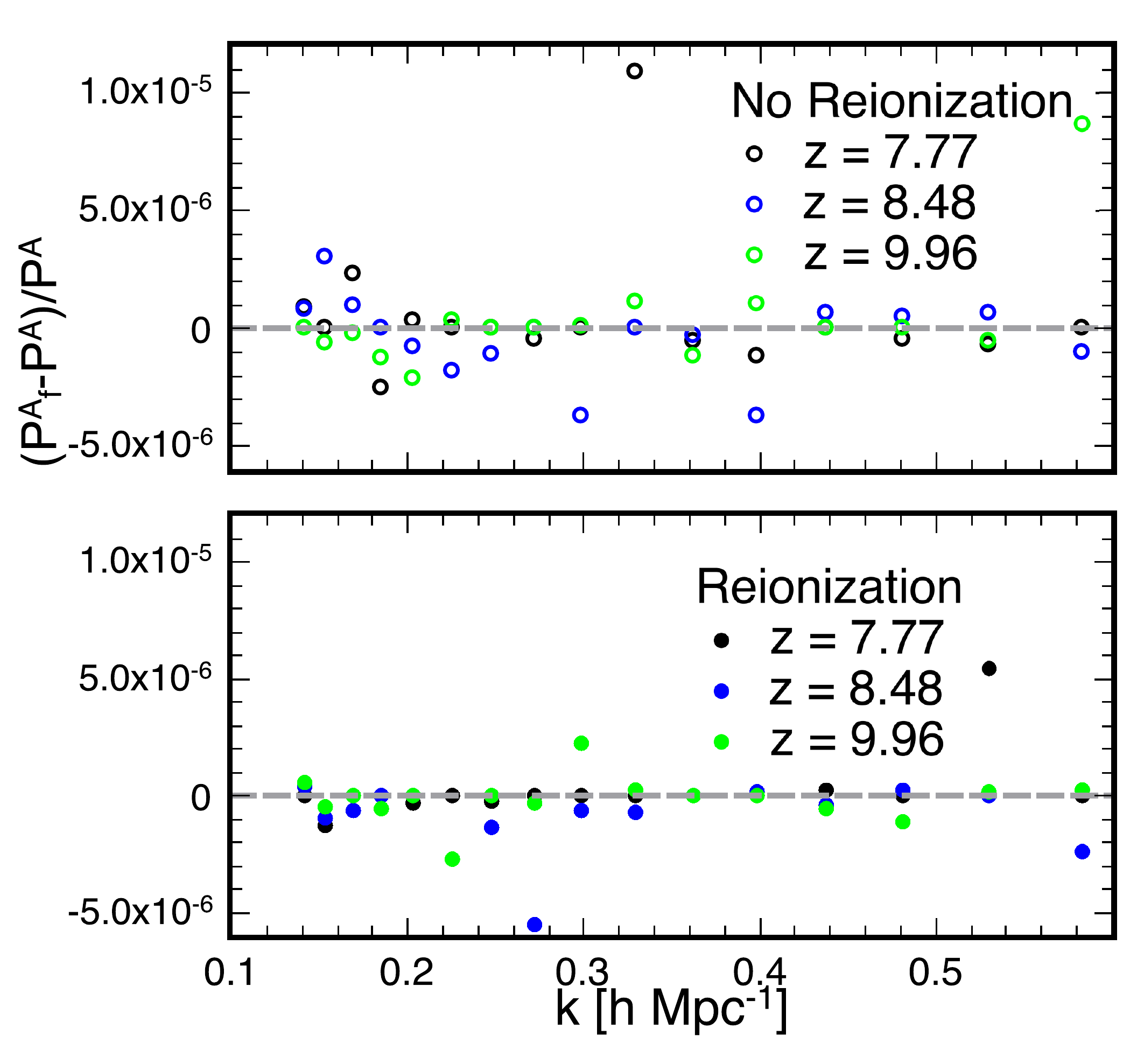} 
\caption{The impact of foreground contamination. We show the fractional change $(P^{\rm A}_{\rm f}-P^{\rm A})/P^{\rm A}$ vs the wavenumber $k$, in terms of $P^{\rm A}_{\rm f}$, the dipole when foregrounds are applied to the 21~cm and CO intensity maps, and $P^{\rm A}$, the dipole with complete removal of foregrounds. We show the results for the hypothetical no-reionization scenario (top) and the fiducial reionization model (bottom) at three redshifts $z=7.77$ (black), $8.48$ (blue) and $9.96$ (green).}
\label{Foreground}
\end{figure}

\subsection{Effect of foreground contamination}

To generate the mock foregrounds for redshifted 21~cm and CO line, we employ the code {\tt Cora}\footnote{\url{https://github.com/radiocosmology/cora}}, which uses the technique described in \cite{Shaw2014All}. This code considers four sources of foregrounds --- Galactic synchrotron, Galactic free-free, extragalactic diffuse free-free and extragalactic point sources. At a given observed frequency, it generates random numbers with Gaussian distribution as real and imaginary parts of the foreground in the 2D Fourier space, which satisfies the analytical form of the foreground power spectrum in \cite{2005ApJ...625..575S}, and then generates a 2D foreground map in the configuration space at a given frequency by the inverse Fourier transformation. 
To create a 3D foreground data cube, we generate with {\tt Cora} the map of 2D slice perpendicular to the LOS at each observed frequency for the 21~cm or CO line emitted from the comoving positions of that 2D slice, and then combine these 2D maps to form a 3D data cube. 
In this paper, we only consider unpolarized foregrounds, and neglect the contribution of other molecular line emissions to the CO foreground \citep{gong2011probing}. We also assume that the foregrounds are completely uncorrelated with the 21~cm and CO signals from the EOR. 

In Fig.~\ref{Foreground} we compare the dipoles with foreground-contaminated maps and with foreground-free maps.  
For both inside-out reionization and the hypothetical no-reionization scenario, we find that the fractional difference between the dipoles with and without foreground is less than one part in 100,000. In other words, the H{\small I}-CO dipole is robust against foregrounds, and thus can be measured directly from the observed, foreground-contaminated, data. This advantage makes the dipole analysis a clean method to extract the information from the 21~cm and CO maps during the EOR. 

This result can be explained by the fact that both 21~cm and CO foregrounds are caused by the same set of radio sources and therefore depend on the observed frequency with exactly the same power law, even though their strengths are orders of magnitude different. 
The foreground on a 3D data cube is generated by mapping the observed frequency of foreground to the corresponding comoving distance and cosmic time on the lightcone. Consequently, the 21~cm and CO foregrounds appear to ``co-evolve'' at the same pace on the lightcone. The evolution effect of antisymmetric cross-correlation for such two ``co-evolving'' foreground fields is negligible, and thus has no impact on the H{\small I}-CO dipole.

\subsection{Model-dependence and independence}
\label{subsec:model-indep}

\begin{figure}
\centering
\includegraphics[width=0.9\columnwidth]{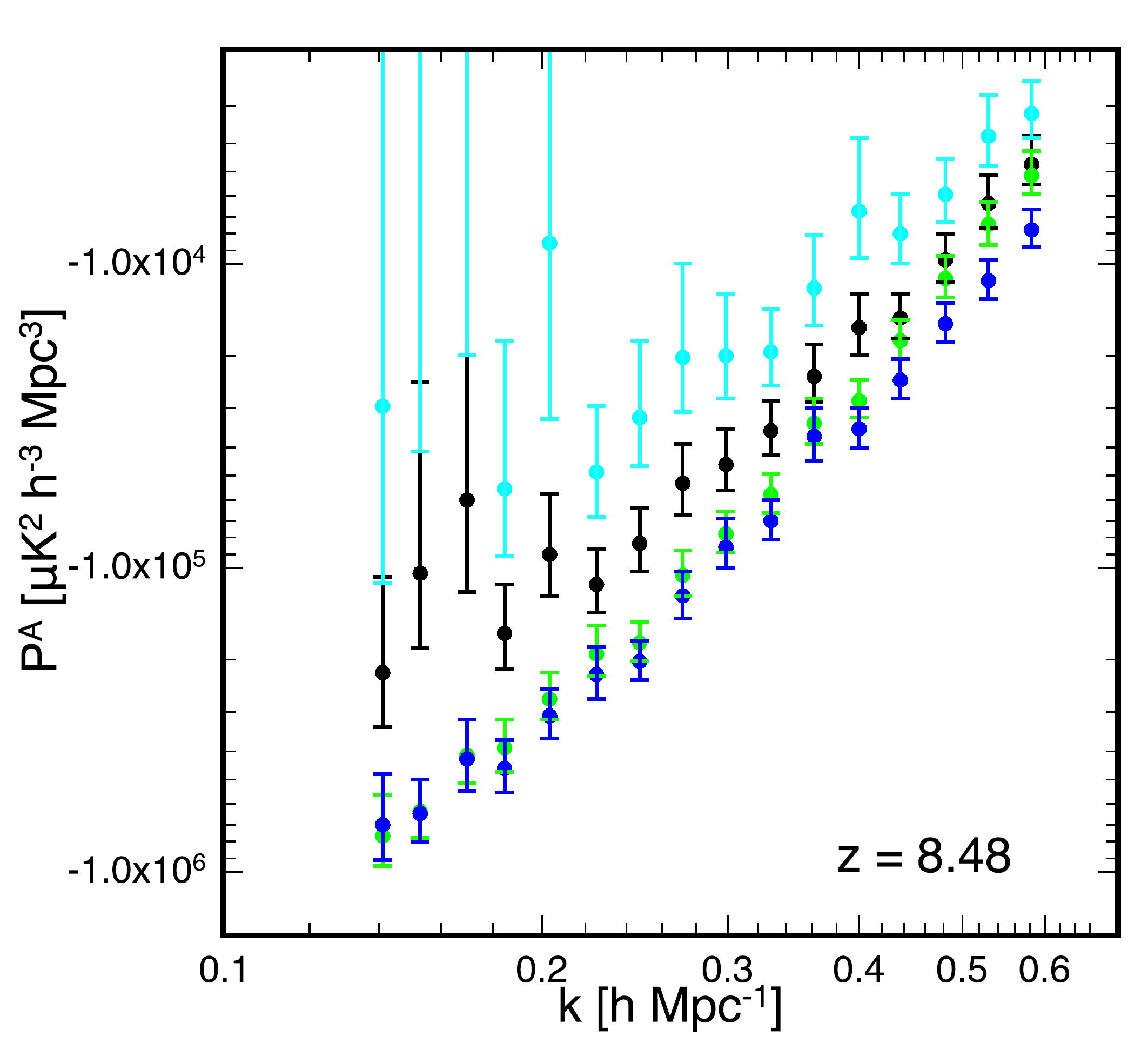}
\caption{The dipole of the H{\small I}-CO cross-power spectrum $P^{\rm A}$ vs wavenumber $k$ for different reionization models (as marked in the legend of Fig.~\ref{AstrophysParameters_globalhistory}) at the fixed redshift $z = 8.48$ (corresponding to $\bar{x}_{\rm HI} = 0.50$ in our fiducial model). The error bars are 1$\sigma$ standard deviation for cosmic variance corresponding to the simulation volume of 100 realizations.}
\label{AstrophysParameters_samez}
\end{figure}

We now consider a wider range of astrophysical parameters in the EOR model, as listed in the legend of Fig.~\ref{AstrophysParameters_globalhistory}. Fig.~\ref{AstrophysParameters_samez} shows that the H{\small I}-CO dipole at the fixed redshift of bandwidth center is very model-dependent.\footnote{Nevertheless, note that varying $R_{\rm mfp}$ can only change the dipole with the fractional difference $\lesssim 20\%$, as we show in the Appendix \ref{app:Rmfp}. Therefore, these EOR models do not involve the variation of $R_{\rm mfp}$.} Thus one might use this model-dependence as the basis of constraining reionization model parameters with the H{\small I}-CO dipole. 

\begin{figure*}
\centering
\includegraphics[width=0.9\columnwidth]{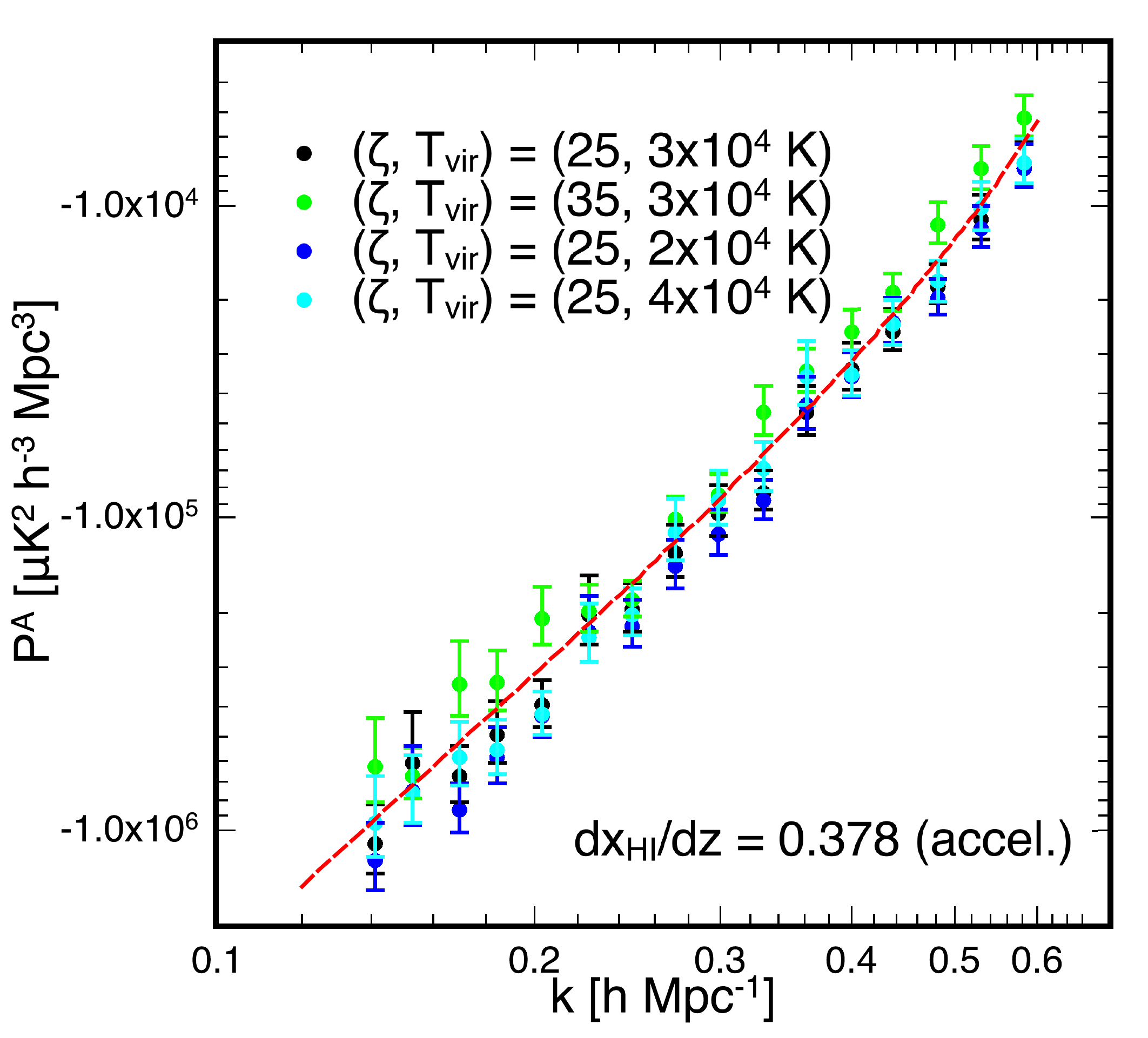}
\includegraphics[width=0.9\columnwidth]{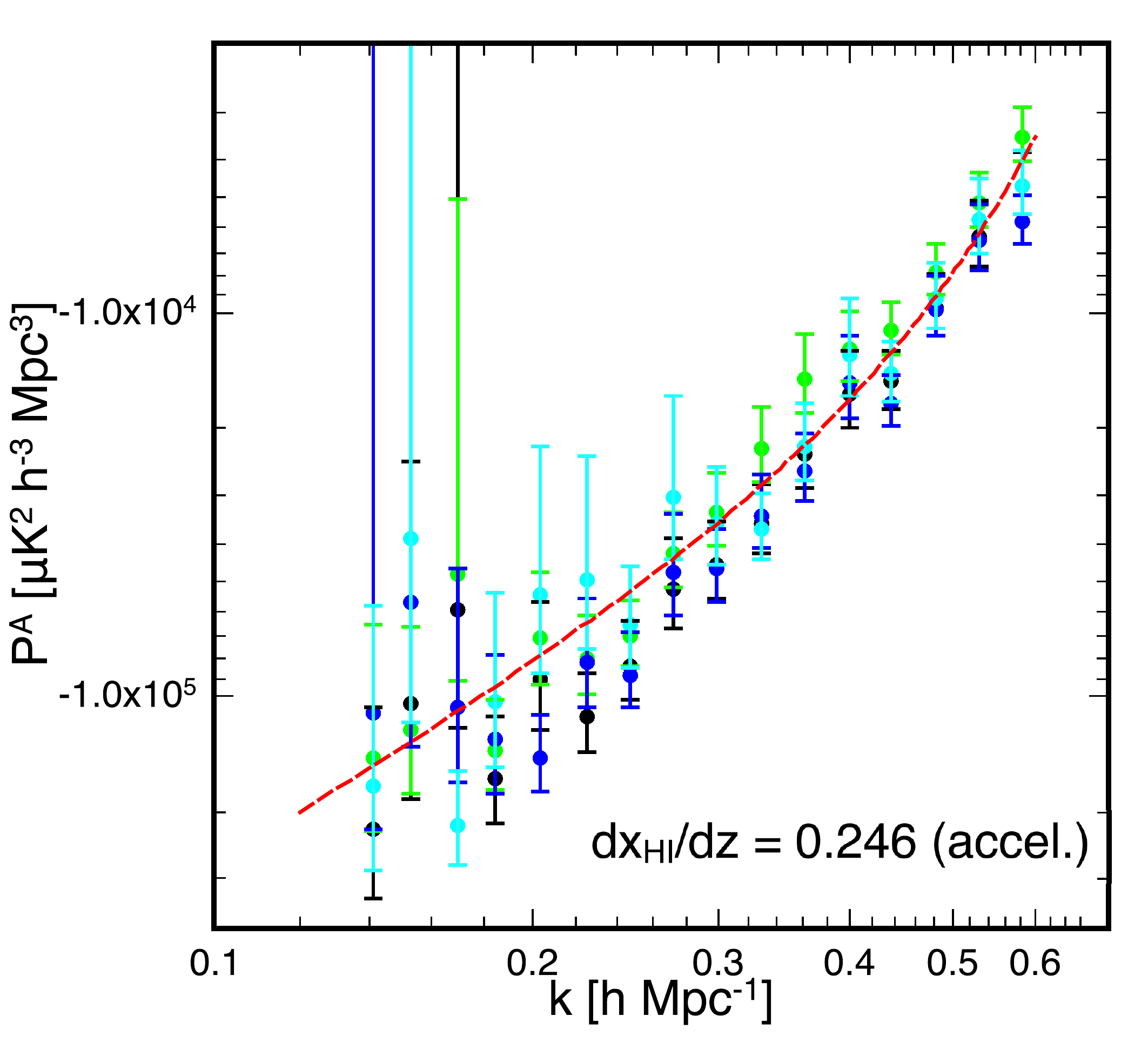}

\includegraphics[width=0.9\columnwidth]{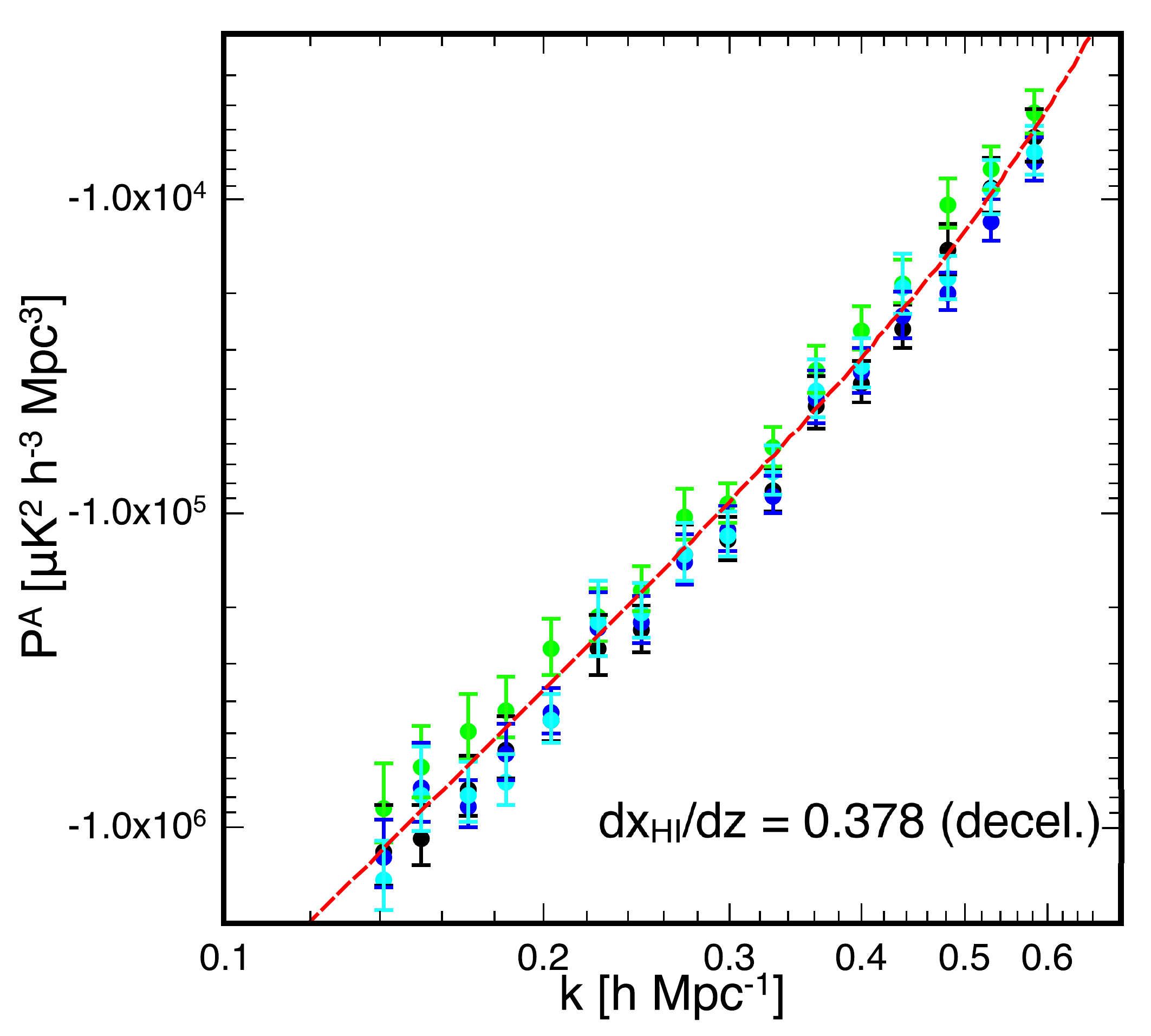}
\includegraphics[width=0.9\columnwidth]{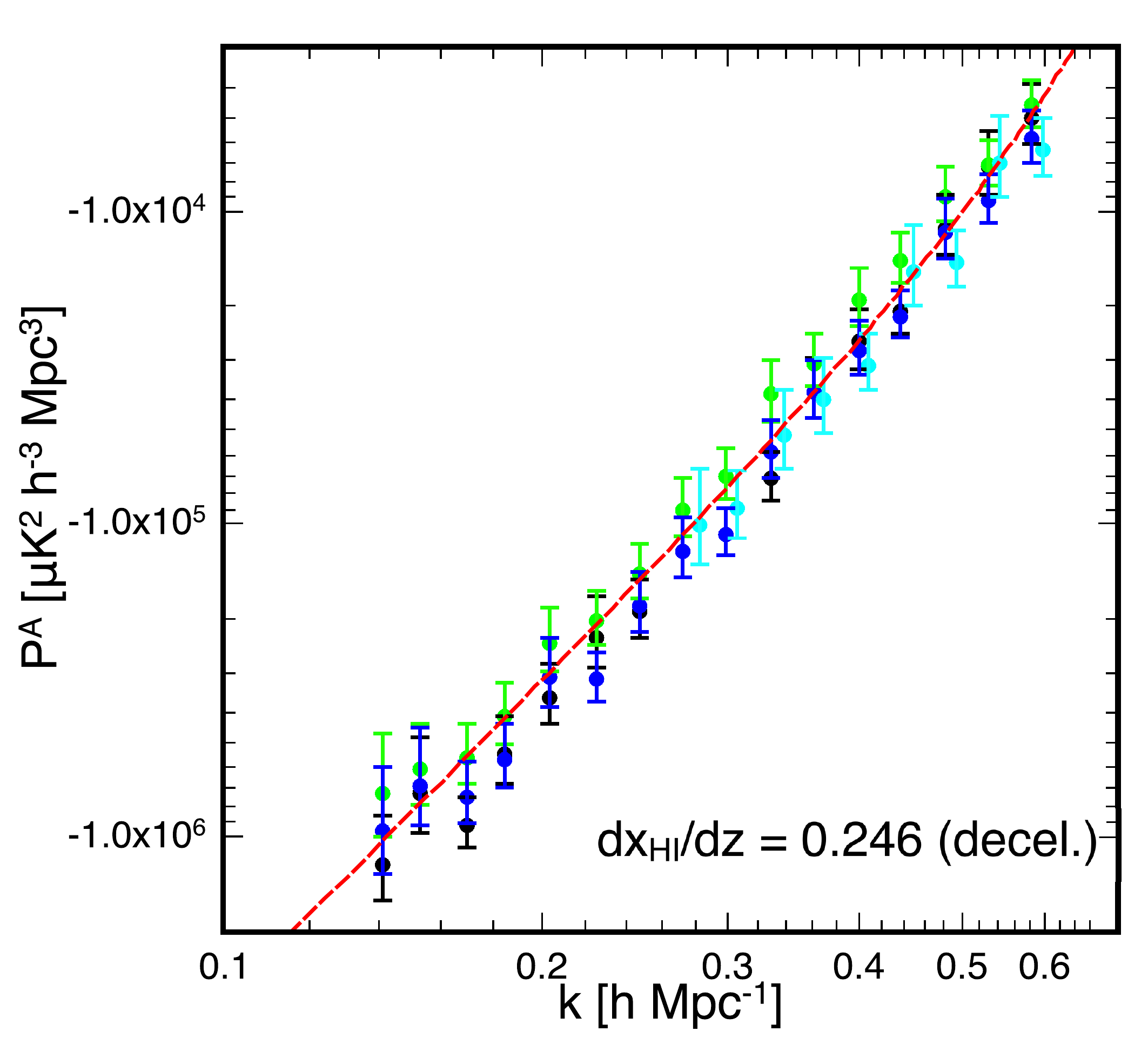}
\caption{The same as Fig.~\ref{AstrophysParameters_samez} but at the fixed rate $d\bar{x}_{\rm HI}/dz = 0.378$ (top left) and $0.246$ (top right) in the acceleration stage and at the fixed $d\bar{x}_{\rm HI}/dz = 0.378$ (bottom left) and $0.246$ (bottom right) in the deceleration stage, respectively. We fit the dipole data of all models (dots) to a modified power law (red dashed lines). }
\label{AstrophysParameters_samedxdz}
\end{figure*}

Fig.~\ref{AstrophysParameters_globalhistory} shows that the progressing of reionization generically undergoes an acceleration stage ($d^2\bar{x}_{\rm HI}/dz^2 < 0$) and a deceleration stage ($d^2\bar{x}_{\rm HI}/dz^2 > 0$). Although the dipole is model-dependent at a fixed $z$, if we compare the dipole for different reionization parameters at the same speed $d\bar{x}_{\rm HI}/dz = 0.378$/$0.246$ of bandwidth center in the acceleration stage, which correspond to $\bar{x}_{\rm HI} = 0.25$/$0.50$ in our fiducial model, respectively, we find in Fig.~\ref{AstrophysParameters_samedxdz} that the dipoles for different models agree with each other, with the scatter within 1$\sigma$ cosmic variance corresponding to the simulation volume of 100 realizations. We find the similar model-independence in the deceleration stage, too. This implies that the H{\small I}-CO dipole is to leading order determined by the speed of cosmic reionization $d\bar{x}_{\rm HI}/dz$ of the bandwidth center, regardless of reionization parameters.
This can be understood because the H{\small I}-CO dipole is dominated by the evolution effect due to cosmic reionization, which is characterized by the difference of ionization level between the front- and back-end on the lightcone, or $d\bar{x}_{\rm HI}/dz$ to leading order. 

\begin{table*}
    \centering
    \caption{Best-fit coefficients of the H{\small I}-CO dipole to the ansatz $P^{\rm A}(k)= -A_{\rm R} (k/k_\ast)^{-n_{\rm R}} \exp{\left[-\beta_{\rm R}(k/k_\ast)^{\alpha_{\rm R}} \right]}$, at the fixed speed $d\bar{x}_{\rm HI}/dz$ for the acceleration stage (``accel.'') and deceleration stage (``decel''), respectively. The slope $n_{\rm R}$ is measured for the range of $k=0.14 - 0.27\,h\,{\rm Mpc}^{-1}$. Here we choose $k_\ast = 1\,h\,{\rm Mpc}^{-1}$. $R^2$ is the coefficient of determination.}
    \begin{tabular}{ccccccc}
    \hline 
  {}  & $d\bar{x}_{\rm HI}/dz$ &  $A_{\rm R}\,[(\mu {\rm K})^2\,h^{-3}\,{\rm Mpc}^3]$  & $n_{\rm R}$ & $\beta_{\rm R}$ & $\alpha_{\rm R}$ & $R^2$ \\
        \hline
\multirow{3}{*}{\rotatebox[origin=c]{90}{accel.}} & $0.378$  &  $(2.539\pm 0.809)\times 10^3$ & $3.018\pm 0.200$ & $3.295\pm 0.976 $ & $ 2.790 \pm 0.430$ & $ 0.9898$ \\
{} & $0.312$  &  $(4.163\pm 2.010)\times 10^3$ & $2.459\pm 0.308$ & $3.877\pm 1.084 $ & $ 2.433 \pm 0.391$ & $ 0.9734$ \\
{} & $0.246$  &  $(6.028\pm 4.584)\times 10^3$ & $1.668\pm 0.500$ & $5.083\pm 1.298 $ & $ 2.513 \pm 0.368$ & $ 0.9542$ \\
    \hline
\multirow{3}{*}{\rotatebox[origin=c]{90}{decel.}} & $0.378$ &  $(1.849\pm 0.624)\times 10^3$ & $3.298\pm 0.211$ & $3.557\pm 1.738 $ & $ 3.266 \pm 0.725$ & $ 0.9861$ \\
{} & $0.312$  &  $(1.842\pm 0.588)\times 10^3$ & $3.273\pm 0.200$ & $3.212\pm 1.189 $ & $ 2.870 \pm 0.533$ & $ 0.9882$ \\
{} & $0.246$  &  $(1.530\pm 0.700)\times 10^3$ & $3.334\pm 0.286$ & $2.729\pm 1.152 $ & $ 2.665 \pm 0.599$ & $ 0.9827$ \\
    \hline 
    \end{tabular}
    \label{tab:coeff2}
\end{table*}

\begin{figure}
\centering
\includegraphics[width=\columnwidth]{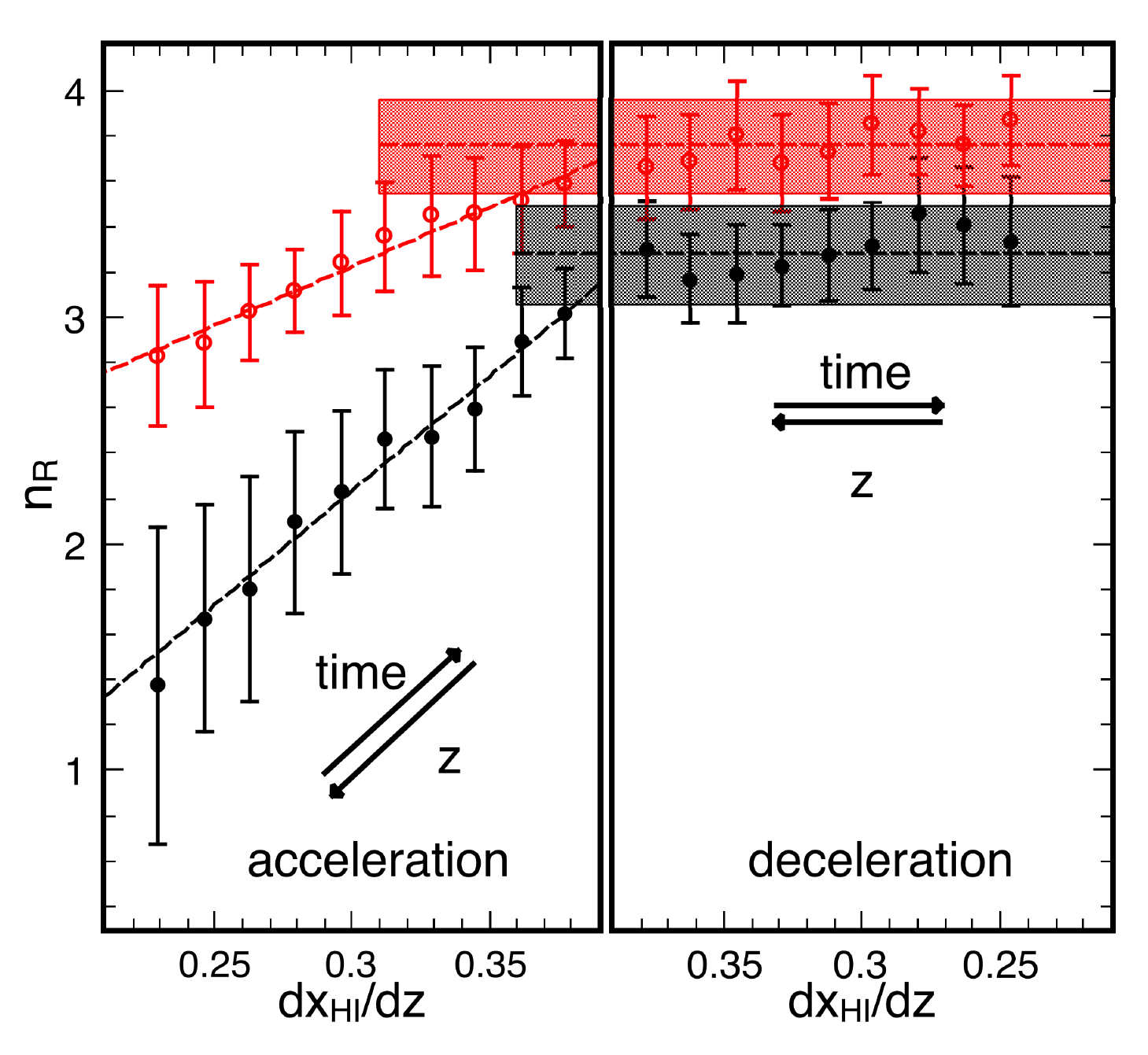}
\caption{The slope of H{\small I}-CO dipole $n_{\rm R}$ vs the speed of reionization $d\bar{x}_{\rm HI}/dz$ during the acceleration (left) and deceleration (right) stage, respectively. We show the slope measured for $k=0.14 - 0.27\,h\,{\rm Mpc}^{-1}$ (black solid dots) and for $k=0.20 - 0.40\,h\,{\rm Mpc}^{-1}$ (red open dots). We fit the data to a linear relation (dashed lines) in the acceleration stage and a constant (dashed lines with the shaded regions representing the 1$\sigma$ error) in the deceleration stage. 
The arrows show the direction of time flow or increasing the redshift $z$.}
\label{nRvsdx/dz}
\end{figure}

At each fixed $d\bar{x}_{\rm HI}/dz$ (during either acceleration or deceleration stage) , we fit the H{\small I}-CO dipole of all models to the modified power law (Eq.~\ref{eqn:dipole_R}) 
with best-fit coefficients listed in Table~\ref{tab:coeff2}. We find that among these four coefficients, $n_{\rm R}$ has the smallest error bars ($\lesssim 15\%$ in most cases) in fitting all model data, which indicates that the slope at large scales is most robust against the variation of models. 

Furthermore, Fig.~\ref{nRvsdx/dz} shows that the slope of H{\small I}-CO dipole increases with the speed $d\bar{x}_{\rm HI}/dz$ with an approximately linear relation, 
$n_{\rm R} = a_1\,(d\bar{x}_{\rm HI}/dz)+a_2$, 
until $n_{\rm R}$ reaches a maximum and levels out near the end of reionization. The linear relation can be explained by the fact that the larger speed $d\bar{x}_{\rm HI}/dz$ can result in larger magnitude of dipole at large scales while having less effect at small scales, which makes the H{\small I}-CO dipole steeper in $k$ and increases the slope $n_{\rm R}$. 
On the other hand, if the slope is measured for a smaller scale, then Fig.~\ref{nRvsdx/dz} shows that the slope reaches the maximum at earlier time. (We define this moment to be when $n_{\rm R}$ overlaps with the 1$\sigma$ region of the constant $n_{\rm R}$ in the deceleration stage.) This implies that the slope reaches the maximum when the scale for which the slope is measured is below the characteristic scale of H{\small II} regions; thereafter the changes in the speed of reionization, which mostly affect the dipole on scales above the characteristic scale of H{\small II} regions, do not considerably affect the scale-dependence of the dipole on scales beneath. 

The linear relation between the slope of H{\small I}-CO dipole at large scales and the speed of reionization can be exploited to infer $d\bar{x}_{\rm HI}/dz$ from the $n_{\rm R}$ measurement in a manner independent of reionization parameters. In this sense, the H{\small I}-CO dipole is a smoking-gun probe for the speed of reionization, so we term it a ``standard speedometer'' for cosmic reionization. 
However, this approach is only valid before $n_{\rm R}$ levels out at the time that depends on the scale of interest. For example, if the range of scale $k=0.14 - 0.27\,h\,{\rm Mpc}^{-1}$ is considered, then the effective range for standard speedometer is $0.25 \lesssim \bar{x}_{\rm HI} \lesssim 0.52$ for all models considered herein. The upper-bound is because at even earlier time, the dipole is too small to be distinguished from the cosmic variance. 

\subsection{Caveats}
Other effects may contribute to the asymmetry as well, but they are suppressed at high redshifts. 
First, the field of view for radio interferometric observations corresponds to only a few hundred comoving Mpc at the EOR, at which scale the relativistic distortions \citep{2009JCAP...11..026M,2013MNRAS.434.3008C,Bonvin_2014,2014PhRvD..89h3535B,Bonvin2015Optimising} are not important. Secondly, the wide-angle effect \citep{Gaztanaga_2017} is negligible for this field of view, and the distant-observer assumption holds well at the high redshifts. Also, the large-angle effect \citep{Gaztanaga_2017} is irrelevant because different choices of angle used to measure the antisymmetry do not make difference under this assumption. Lastly, gravitational lensing \citep{2014PhRvD..89h3535B,PhysRevLett.124.031101} of the 21~cm line is a secondary, higher-order, effect. 


\section{Summary and Conclusions}
\label{sec:conclusions}
In this paper, we propose a new intensity mapping analysis method, the antisymmetric cross-correlation between the H{\small I} 21~cm line and the CO 2.61~mm line intensity maps from the EOR as a new probe of cosmic reionization. 
From the observational point of view, the most interesting advantage of this new analysis is that this signal is unbiased by foregrounds and thus can be measured directly from the foreground-contaminated data. It arises because the statistical fluctuations of the 21~cm field have much more rapid evolution in time than the CO(1-0) line field, and therefore the H{\small I}-CO antisymmetric cross-correlation contains additional information of the progressing of cosmic reionization, complementary to the symmetric component of cross-correlation. The sign of this signal can generically tell whether inside-out reionization happens during some time interval, regardless of the detail of reionization model. 

We use the dipole of H{\small I}-CO cross-power spectrum as the estimator of the antisymmetric cross-correlation, and find that the H{\small I}-CO dipoles for different astrophysical parameters at the time corresponding to the same $d\bar{x}_{\rm HI}/dz$ agree with each other within 1$\sigma$ error of sampling variance in our simulations. This means that the dipole is to leading order sensitive to the speed of reionization $d\bar{x}_{\rm HI}/dz$, regardless of reionization parameters, because the antisymmetric H{\small I}-CO cross-correlation is dominated by the evolution effect due to cosmic reionization. In particular, the slope of the H{\small I}-CO dipole at large scales, which we find is the least sensitive to the model uncertainties, is linear to $d\bar{x}_{\rm HI}/dz$, until the slope levels out near the end of reionization. The H{\small I}-CO dipole, therefore, may be a smoking-gun probe for the speed of reionization, or a standard speedometer for cosmic reionization. 

Standard speedometer is of important astrophysical application --- the global reionization history $\bar{x}_{\rm HI}(z)$, however relative to the value when the slope just begins to level out, can be reconstructed by integration of $d\bar{x}_{\rm HI}/dz$. This approach of global history reconstruction, albeit only valid from the midpoint to near the completion of reionization, is independent of reionization parameters, and unbiased by foregrounds. Measurements of the H{\small I}-CO dipole will motivate the greater synergy between 21~cm observations using radio interferometers, e.g. the low frequency array of the Square Kilometre Array (SKA) \citep{2015aska.confE...1K}, and CO observations using single dish arrays, e.g. the update of the CO Mapping Array Pathfinder (COMAP) \citep{2016AAS...22742606C,2016ApJ...817..169L}. In a companion paper (Tan et al. in prep.), we will explore the detectability of the H{\small I}-CO dipole and the prospect of global reionization history reconstruction using standard speedometer with futuristic radio observations. 

Note, however, that our demonstration is made only within the framework of the excursion set model of reionization, in which the reionization modelling is parameterized by three parameters: $\zeta$ (the ionizing efficiency), $T_{{\rm vir}}$ (the minimum virial temperature of halos that host ionizing sources), and $R_{\rm mfp}$ (the maximum mean free path of ionizing photons). Thus it is possible that the relation between the slope of the H{\small I}-CO dipole and the speed of reionization depends on different frameworks of reionization modelling. In that case, the inference of the global reionization history from the standard speedometer must take into account the possible systematic error due to theoretical uncertainties of different reionization modelling.

In principle, the H{\small I} 21~cm map can be cross-correlated with other tracers of cosmological density fluctuations from the EOR, e.g.\ [C{\small II}] \citep{2012ApJ...745...49G} or other molecular line intensity maps, or high-redshift galaxy surveys \citep{2007ApJ...660.1030F,2009ApJ...690..252L}, if possible. As the evolution effect due to cosmic reionization dominates over that of the density fluctuations, we expect that the dipole of the cross-power spectrum between the 21~cm line and a generic probe of density fluctuations from the EOR can be a {\it class} of new probes for cosmic reionization that have similar features.

{\it Note.--} While \cite{2020PhRvD.102d3519S} was published earlier than this paper, the proposal of using antisymmetric cross-correlation between the 21~cm and CO line intensity maps as a new probe of cosmic reionization was originally presented with main results by Y.M.~in his talk at the 2019 LIM conference held at the CCA, NYC, as acknowledged by \cite{2020PhRvD.102d3519S} themselves. Our paper has essentially least overlapping with \cite{2020PhRvD.102d3519S} but this proposal. \cite{2020PhRvD.102d3519S} focused on the antisymmetric part of angular cross-power spectrum and its application to a {\it special} reionization model, instead of generic features of this new signal which are the focus of our paper. 

\section*{Acknowledgements}
This work is supported by the National Key R\&D Program of China (Grant No.2018YFA0404502, 2017YFB0203302),
and the National Natural Science Foundation of China (NSFC Grant No.11673014, 11761141012, 11821303).  
YM was also supported in part by the Chinese National Thousand Youth Talents Program. 
We are grateful to Andrea Ferrara, Ben Wandelt, Le Zhang and Pengjie Zhang for comments and discussions. The simulations in this work were ran at the Venus cluster at the Tsinghua University.

\begin{appendix}

\section{Variance} 
\label{app:var}
We derive the variance of antisymmetric cross-power spectrum (Eq.~\ref{eq:var}) here. We start from the definition in the discretized form, 
$V_{\rm tot} P_{ij}^{\rm I}(\textbf{k}) = \frac{1}{2i} \left[\widetilde{\delta_i}(\textbf{k})\widetilde{\delta_j}^\ast(\textbf{k})- \widetilde{\delta_i}^\ast(\textbf{k})\widetilde{\delta_j}(\textbf{k})\right]$, 
where $V_{\rm tot} \delta_{\textbf{k},\textbf{k}'} \to (2\pi)^3\delta_{\rm D}^{(3)}(\textbf{k}-\textbf{k}')$ in the limit $V_{\rm tot}\to\infty$. This yields
\begin{widetext}
\begin{align}
   & \phantom{=} \left< P_{ij}^{\rm I}(\textbf{k}) P_{ij}^{\rm I}(\textbf{k}') \right> = -\frac{1}{4\,V_{\rm tot}^2}\, \left< \left[\widetilde{\delta_i}(\textbf{k})\widetilde{\delta_j}^\ast(\textbf{k})- \widetilde{\delta_i}^\ast(\textbf{k})\widetilde{\delta_j}(\textbf{k})\right] \left[\widetilde{\delta_i}(\textbf{k}')\widetilde{\delta_j}^\ast(\textbf{k}')- \widetilde{\delta_i}^\ast(\textbf{k}')\widetilde{\delta_j}(\textbf{k}')\right] \right> \notag \\
   & = -\frac{1}{4\,V_{\rm tot}^2}\,\left\{ \left< \widetilde{\delta_i}(\textbf{k}) \widetilde{\delta_j}^\ast(\textbf{k}) \widetilde{\delta_i}(\textbf{k}')\widetilde{\delta_j}^\ast(\textbf{k}') \right> - \left< \widetilde{\delta_i}(\textbf{k}) \widetilde{\delta_j}^\ast(\textbf{k}) \widetilde{\delta_i}^\ast(\textbf{k}')\widetilde{\delta_j}(\textbf{k}') \right> + {\rm c.c.}\right\} \notag \\
   & = -\frac{1}{4\,V_{\rm tot}^2}\,\left\{ \left< \widetilde{\delta_i}(\textbf{k}) \widetilde{\delta_j}^\ast(\textbf{k}) \right>\left< \widetilde{\delta_i}(\textbf{k}')\widetilde{\delta_j}^\ast(\textbf{k}') \right> + \left< \widetilde{\delta_i}(\textbf{k})\widetilde{\delta_i}(\textbf{k}')\right>\left< \widetilde{\delta_j}^\ast(\textbf{k}) \widetilde{\delta_j}^\ast(\textbf{k}') \right> + \left< \widetilde{\delta_i}(\textbf{k}) \widetilde{\delta_j}^\ast(\textbf{k}')\right>\left< \widetilde{\delta_j}^\ast(\textbf{k}) \widetilde{\delta_i}(\textbf{k}') \right> \right.\notag \\
   & \phantom{=} \left. - \left< \widetilde{\delta_i}(\textbf{k}) \widetilde{\delta_j}^\ast(\textbf{k})\right>\left< \widetilde{\delta_i}^\ast(\textbf{k}')\widetilde{\delta_j}(\textbf{k}') \right> - \left< \widetilde{\delta_i}(\textbf{k})\widetilde{\delta_i}^\ast(\textbf{k}')\right>\left< \widetilde{\delta_j}^\ast(\textbf{k}) \widetilde{\delta_j}(\textbf{k}') \right> - \left< \widetilde{\delta_i}(\textbf{k}) \widetilde{\delta_j}(\textbf{k}')\right>\left< \widetilde{\delta_j}^\ast(\textbf{k}) \widetilde{\delta_i}^\ast(\textbf{k}') \right> + {\rm c.c.} \right\} \notag \\
   & = \left< P_{ij}^{\rm I}(\textbf{k})\right> \left<P_{ij}^{\rm I}(\textbf{k}')\right>   + \frac{1}{2}\, \left[\delta_{\textbf{k},\textbf{k}'}-\delta_{\textbf{k},-\textbf{k}'} \right] \left[ P_{i}(k) P_{j}(k) -\left(P_{ij}^{\rm S}(k)\right)^2 + \left(P_{ij}^{\rm I}(\textbf{k}) \right)^2\right] \,.\nonumber
\end{align}
\end{widetext}
Here ``c.c.''~stands for the complex conjugate of the preceding terms. In the third line above, we used the Wick theorem. Since $P_{ij}^{\rm I}(-\textbf{k}) = - P_{ij}^{\rm I}(\textbf{k})$, the $-\textbf{k}$ mode is not independent from the $\textbf{k}$ mode, so we only consider the upper hemisphere in $\textbf{k}$-space, and the variance combines the contribution from both $\textbf{k}$ and $-\textbf{k}$ modes, i.e. $\sigma^2_{P_{ij}^{\rm I}}({\bf k}) = \Bigl[\left< \left(P_{ij}^{\rm I}(\textbf{k})\right)^2 \right> - \left< P_{ij}^{\rm I}(\textbf{k})\right>^2\Bigr] - \Bigl[ \left< P_{ij}^{\rm I}(\textbf{k}) P_{ij}^{\rm I}(-\textbf{k}) \right> -  \left< P_{ij}^{\rm I}(\textbf{k})\right> \left<P_{ij}^{\rm I}(-\textbf{k})\right>\Bigr] $. Therefore
\begin{equation}
\sigma^2_{P_{ij}^{\rm I}}({\bf k}) = P_{i}(k) P_{j}(k) -\left(P_{ij}^{\rm S}(k)\right)^2 + \left(P_{ij}^{\rm I}(\textbf{k}) \right)^2\,.\nonumber
\end{equation}
When average the signal over modes with the same $(k,\theta)$, the variance is reduced by a factor of $N(k,\theta)$ (the number of these modes in the upper hemisphere), and thus we find Eq.~(\ref{eq:var}). 

\begin{figure*}
\centering
\includegraphics[width=0.9\columnwidth]{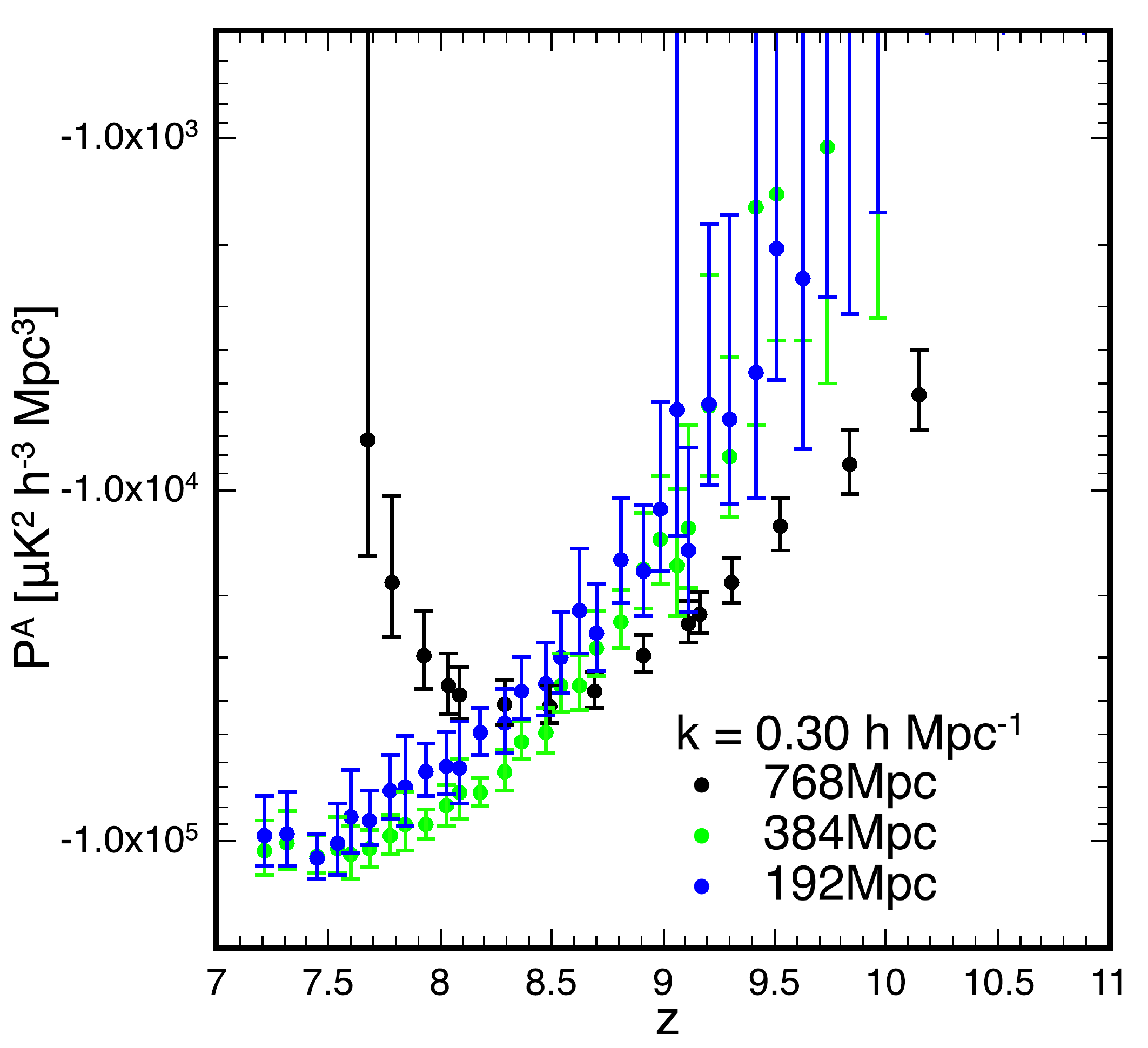}
\includegraphics[width=0.9\columnwidth]{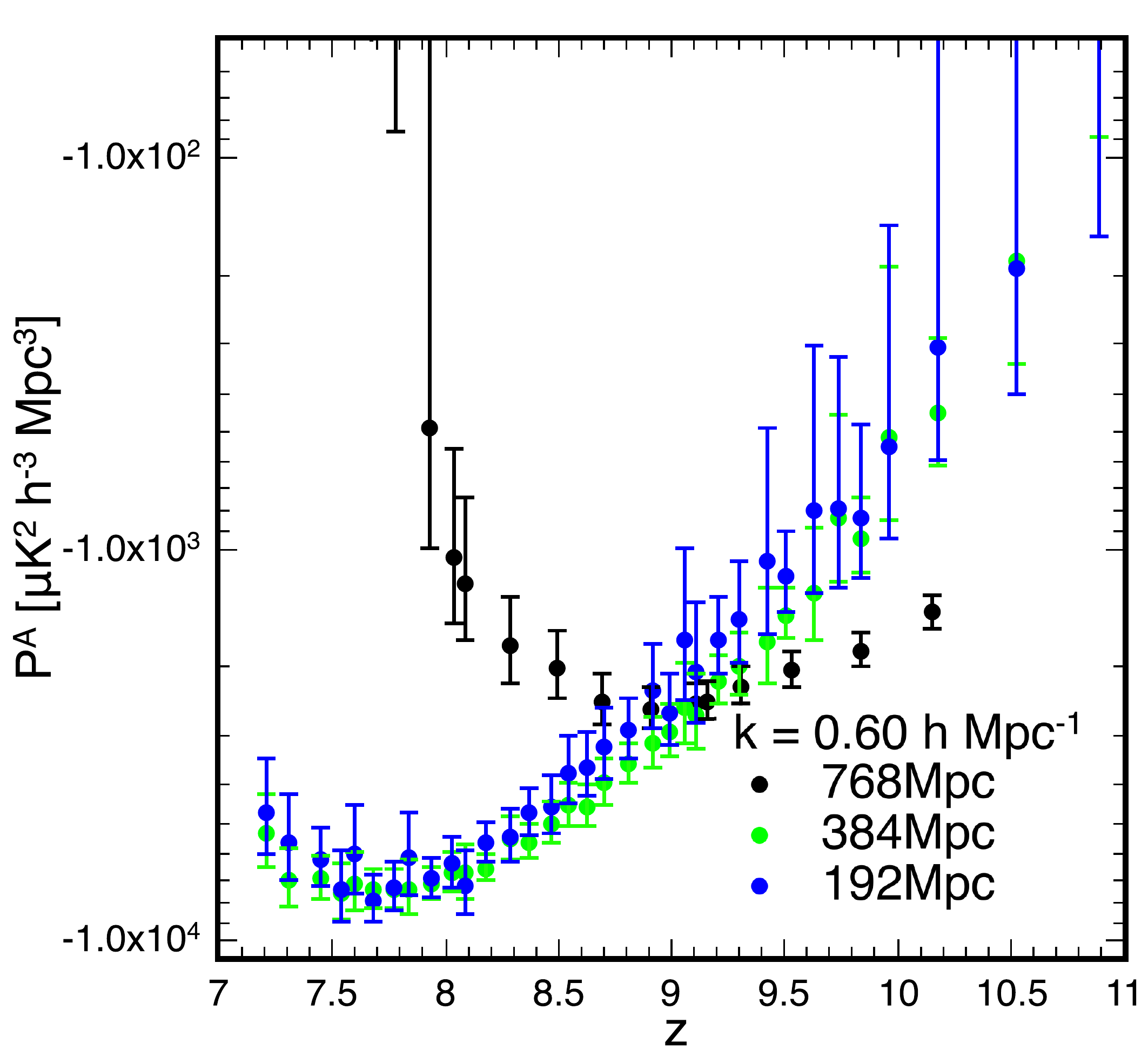}
\caption{Effect of redshift-bin size and finite simulation box. We show the dipole of the H{\small I}-CO cross-power spectrum $P^{\rm A}$ vs central redshift $z$ for our fiducial reionization model at the fixed wavenumber $k=0.30\,h\,{\rm Mpc}^{-1}$ (left) and $k=0.60\,h\,{\rm Mpc}^{-1}$ (right). The results are obtained by cross-correlating the fields in the full simulation volume of 768 comoving Mpc on each side (black), in the inner cubic region of $384$ comoving Mpc on each side away from the boundary of box (green), or shrinking the fields to even smaller size of $192$ comoving Mpc on each side (blue), respectively. The error bars are 1$\sigma$ standard deviation for cosmic variance corresponding to the simulation volume of 100 realizations.}
\label{Boxsize}
\end{figure*}

\section{Effect of redshift-bin size and finite simulation box}
\label{app:conv}
Not only does the antisymmetric cross-power spectrum depend on the central redshift of a volume, but also it depends on the frequency bandwidth or the size of redshift-bin, corresponding to the physical size of the correlated fields. While large redshift-bin size enlarges the accessible scales of dipole, it may also smooth out small-scale asymmetry across different cosmic time, if any. Another issue is the effect of the periodic boundary condition on a finite simulation volume which may also cause systematic errors to the theoretical computation of the dipole from simulations. 

To test these effects, we consider three scenarios for the cross-correlation --- (i) using the full simulation volume (768 comoving Mpc on each side), (ii) using the inner cubic region of 384 comoving Mpc on each side away from the boundary  of box, and (iii) further shrinking the cross-correlated volume to 192 comoving Mpc on each side. In Fig.~\ref{Boxsize}, the results of $(384\,{\rm Mpc})^3$ volume agree with those of $(192\,{\rm Mpc})^3$ volume, while the results of full volume significantly over-/under-estimate the magnitude of dipole at the earlier/late time. As such, the redshift-bin size corresponding to 384 comoving Mpc is a trade-off between accessing the large-scale modes and keeping small-scale asymmetry, which also avoids the finite box effect in our simulations. Our analysis in this paper, therefore, used the data cube constructed from the inner cubic region of 384 comoving Mpc on each side away from the boundary of simulation box.   

\section{Effect of maximum mean free path of ionizing photons}
\label{app:Rmfp}
\begin{figure}
\centering
\includegraphics[width=\columnwidth]{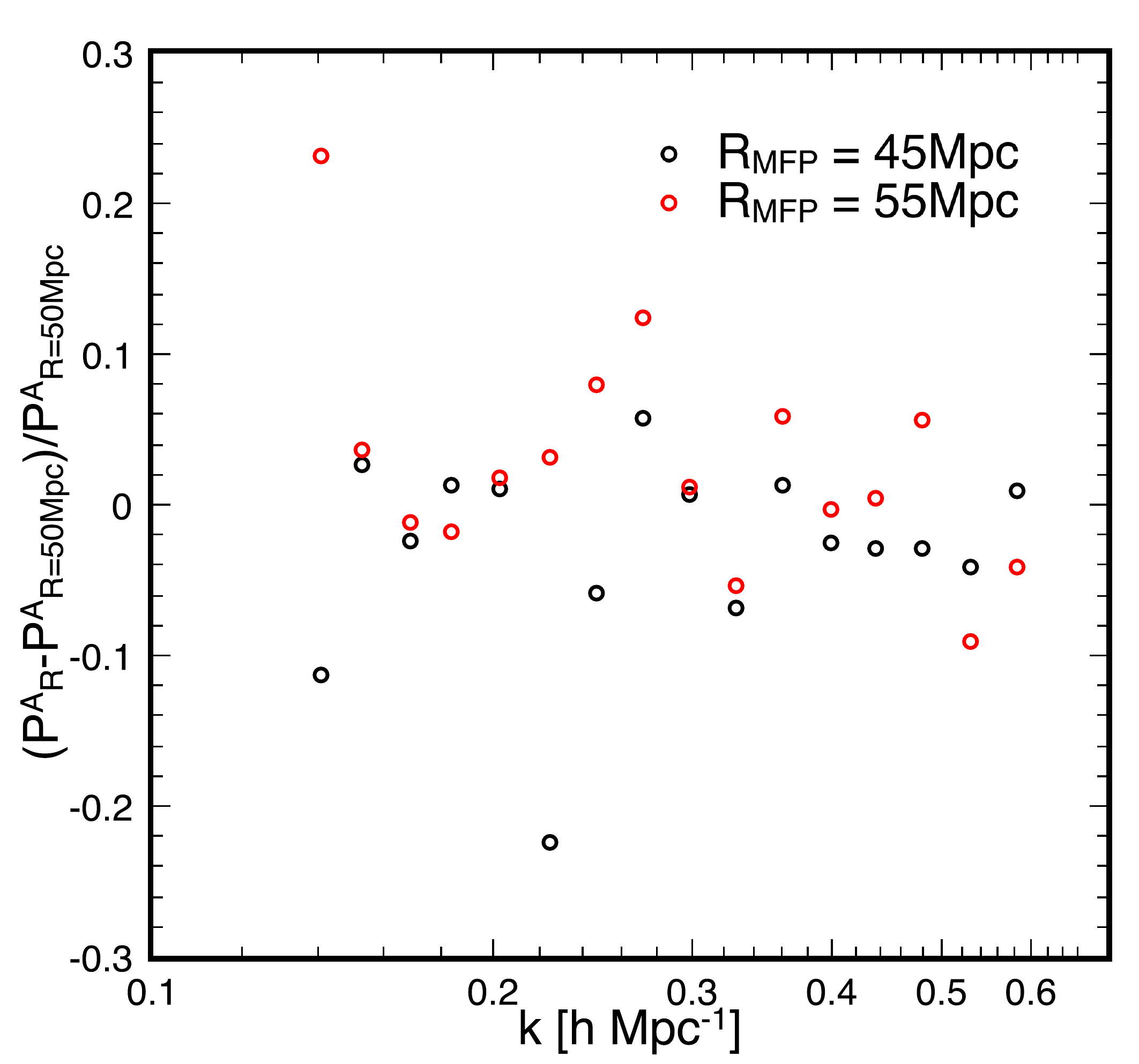}
\caption{The effect of maximum mean free path of ionizing photons ($R_{\mathrm{mfp}}$). We show the fractional change of the dipole $P^{\rm A}_{\rm R}$ with $R_{\mathrm{mfp}}=45/55\,{\rm Mpc}$ (black/red) with respect to the fiducial model with $R_{\mathrm{mfp}} = 50\,{\rm Mpc}$, as a function of the wavenumber $k$ at the fixed redshift of bandwidth center $z = 8.48$ in one realization of simulation. Two other reionization parameters are fixed here, $\zeta = 25$ and $T_{\rm vir} = 3\times10^4\,\rm K$.} 
\label{Rmfpfd}
\end{figure}

In Fig.~\ref{Rmfpfd}, we find that within a reasonable range of parameter variation, the H{\small I}-CO dipole is very weakly dependent of $R_{\rm mfp}$ (maximum mean free path of ionizing photons), with the fractional change $\lesssim 20\%$, with respect to our fiducial EOR model at the fixed central redshift. In comparison, varying other two parameters, $\zeta$ and $T_{\rm vir}$, can result in much more significant difference at the fixed central redshift, as shown in Fig.~\ref{AstrophysParameters_samez}. This is actually expected, because the simulation of H{\small I} field is not affected much as long as $R_{\rm mfp}$, which determines the maximal value of the photon mean free path, is large enough so that the photon traveling is not capped artificially. We thus fix the value of $R_{\rm mfp}$, due to its weak effect, when investigating the model-dependence of dipole in the rest of this paper.

\end{appendix}

\bibliographystyle{aasjournal}
\bibliography{antisymref}

\end{document}